\documentclass[aps,prd,preprintnumbers,nofootinbib,superscriptaddress,preprint]{revtex4-1}
\usepackage{hyperref}
\usepackage{graphicx}
\usepackage{amssymb}
\usepackage{amsmath}
\usepackage{color}
\usepackage{grffile}
\usepackage{booktabs}
\usepackage{dcolumn}

\graphicspath{{fig/}}

\newcommand{\beq}{\begin{eqnarray}}
\newcommand{\eeq}{\end{eqnarray}}

\def\phibar{\overline{\phi}}
\def\muhat{\hat{\mu}}
\newcommand{\Kmu}[2]{K^{(#1)}_{#2,\mu}}

\newcommand{\Ldmu}[2]{L^{(#1)}_{#2,\mu}}

\def\Vc{{\mathcal V}}
\def\Ac{{\mathcal A}}

\def\fsl#1{\setbox0=\hbox{$#1$}
   \dimen0=\wd0
   \setbox1=\hbox{/} \dimen1=\wd1
   \ifdim\dimen0>\dimen1
      \rlap{\hbox to \dimen0{\hfil/\hfil}}
      #1
   \else
      \rlap{\hbox to \dimen1{\hfil$#1$\hfil}}
      /
   \fi}

\newcommand{\nn}{\nonumber}
\newcommand{\Sprm}{$S$ parameter }

\newcommand{\PTSprm}{Peskin-Takeuchi $S$ parameter }
\newcommand{\SUN}[1]{$SU(#1)$}
\newcommand{\SUNV}[1]{$SU(#1)_V$}
\newcommand{\SUNLR}[1]{$SU(#1)_L\times SU(#1)_R$}
\newcommand\supvma{V-A}

\newcommand \ZAEnsembles {\left\{(48, 0.009), (42, 0.012), (42, 0.015), (36, 0.02), (30, 0.03)\right\}}

\newcommand \ZAValue {1.0280(2)}
\newcommand \ZAStatisticalPercentage {0.004\%}
\newcommand \ZASystematicPercentage {0.02\%}

\newcommand \PadeFitLargestChisquare {0.5}
\newcommand \PadeFitSecondLargestChisquare {0.2}

\newcommand \SParameterVPZeroPointZeroZeroNine {0.273(7)}

\newcommand \vpflightensemblesMassRange {0.009 - 0.02}

\newcommand \vpfheavyensemblesMassRange {0.03 - 0.1}

\newcommand \NfEightmfZeroPointZeroZeroNineTSixFourLFourEightPlateauStart {20}

\newcommand \NfEightmfZeroPointZeroZeroNineTSixFourLFourEightPlateauRange {[20, 32]}
\newcommand \NfEightmfZeroPointZeroZeroNineTSixFourLFourEightChisquarePerDof {9.6/9}
\newcommand \NfEightmfZeroPointZeroZeroNineTSixFourLFourEightSParameter {0.2643(42)(5)}

\newcommand \LightEnsemblesWithSystematics {\left\{(48, 0.009), (42, 0.012), (36, 0.015), (36, 0.02), (30, 0.03), (30, 0.04)\right\}}
\newcommand \SingleVolumeEnsembles {\left\{0.009, 0.012, 0.06, 0.08, 0.1\right\}}
\newcommand \MultiVolumeEnsembles {\left\{0.015, 0.02, 0.03, 0.04\right\}}
\newcommand \AllLatticeVolumes {\left\{(64, 48), (56, 42), (48, 36), (40, 30), (32, 24), (24, 18)\right\}}

\newcommand \TimeMomentInfiniteVolumeC {-6.6(1.0)}
\newcommand \TimeMomentInfiniteVolumeChisquareDof {6.2/4}

\newcommand \LatticeSpacingRatioLSDLatKMI {1.29(19)}

\newcommand \lTenrNfEightmfZeroPointZeroZeroNineLFourEightTSixFour {-0.00573(11)(33)}

\newcommand \RhoMassNfEightmfZeroPointZeroZeroNineLFourEightTSixFour {0.2225(24)}
\newcommand \MpiOverMrhoRange {0.63 - 0.77}

\newcommand \SParameterInfiniteVolumeTimeMomentNfEightmfZeroPointZeroZeroNine {0.281(5)(17)}

\renewcommand{\midrule}{\hline}

\begin{document}

\preprint{UTHEP-825, UTCCS-P-181}

\title{The Peskin-Takeuchi $S$ parameter and vector meson decay constants in $N_f=8$ QCD}

\date{\today}

\author{Yasumichi~Aoki}
\affiliation{RIKEN Cener for Computational Sciences (R-CCS),\\ Kobe 650-0047, Japan}

\author{Tatsumi~Aoyama}
\affiliation{Institute for Solid State Physics, University of Tokyo,\\
5-1-5 Kashiwanoha, Kashiwa-shi, Chiba 277-8581, Japan}

\author{Ed~Bennett}
\affiliation{Swansea Academy of Advanced Computing, Swansea University,\\
Bay Campus, Swansea, SA1 8EN, UK}
\affiliation{Centre for Quantum Fields and Gravity, Department of Physics, Swansea University, Singleton Park, Swansea, SA2 8PP, UK}

\author{Toshihide~Maskawa\footnote{Deceased, July 23, 2021.}}
\affiliation{Kobayashi-Maskawa Institute for the Origin of Particles and the Universe, \\
Nagoya University, Nagoya 464-8602, Japan}

\author{Kohtaroh~Miura}
\affiliation{Institute of Pure and Applied Sciences, University of Tsukuba, \\ Tsukuba, Ibaraki 305-8571, Japan}
\affiliation{Institute of Particle and Nuclear Studies,\\
High Energy Accelerator Research Organization (KEK), Tsukuba 305-0801, Japan}

\author{Hiroshi~Ohki}
\affiliation{Department of Physics, Nara Women\'s University,\\
 Nara 630-8506, Japan}

\author{Enrico~Rinaldi}
\affiliation{
Interdisciplinary Theoretical \& Mathematical Science Program, RIKEN (iTHEMS),\\
 2-1 Hirosawa, Wako, Saitama, 351-0198, Japan
}
\affiliation{Quantinuum K.K., Otemachi Financial City Grand Cube 3F \\
1-9-2 Otemachi, Chiyoda-ku, Tokyo, Japan}
\affiliation{Quantinuum, Partnership House, Carlisle Place, London SW1P 1BX, UK}

\author{Akihiro~Shibata}
\affiliation{Computing Research Center, High Energy Accelerator Research Organization (KEK),\\ Tsukuba 305-0801, Japan}

\author{Koichi~Yamawaki}
\affiliation{Kobayashi-Maskawa Institute for the Origin of Particles and the Universe, \\ Nagoya University, Nagoya 464-8602, Japan}

\author{Takeshi~Yamazaki}
\affiliation{Institute of Pure and Applied Sciences, University of Tsukuba, \\ Tsukuba, Ibaraki 305-8571, Japan}
\affiliation{Center for Computational Sciences, University of Tsukuba, \\ Tsukuba, Ibaraki 305-8577, Japan}

\collaboration{LatKMI Collaboration}
\noaffiliation

\begin{abstract}
  We use lattice gauge theory to investigate the \PTSprm in the
  \SUN{3} gauge theory with eight light fundamental fermions (eight-flavor QCD),
  as a candidate for the origin of electroweak symmetry breaking.
  Eight-flavor QCD has been indicated to provide the realization
  of an approximately conformal (walking) technicolor model.
  The fate of the \Sprm in a theory with walking dynamics is of great interest
  with respect to the model building
  for physics beyond the Standard Model (BSM).
  This work develops a new approach to computing the $S$ parameter utilizing the first Time Moment of current correlators,
  in addition to the traditional one using the vacuum polarization.
  The results from both methods are consistent. 
  We evaluate corrections to $S$ for finite-volume (FV) effects
  using our data for multiple lattice volumes
  combined with a FV formula motivated by chiral perturbation theory.
  At the smallest fermion mass, we find the one-doublet contribution to be $S=\SParameterInfiniteVolumeTimeMomentNfEightmfZeroPointZeroZeroNine$,
  which is smaller than in the scaled-up version of 2+1-flavor QCD $\sim 0.35$ (without subtracting the SM contribution).
  In contrast to naive expectations, however, the value of
  $S$ approached at smaller fermion mass is not necessarily
 suppressed when FV corrections are taken into account.
 Using the obtained $S$, 
  the Gasser-Leutwyler parameter $L_{10}^r(M_\rho)$
  results in the similar value to that in real-world QCD.
 Additionally, we examine the vector/axial-vector meson decay constants,
 $F_\rho, F_{a_1}$, using the same datasets.
 Our result for $F_{\rho}/F_{\pi}$ are stable against fermion masses and consistent with
  the estimate of Kawarabayashi-Suzuki-Riazuddin-Fayyazuddin relations (KSRF),
  based on which the coupling {$g_{\rho\pi\pi}$} between
 the rho meson and two pions is estimated
  to have a similar value to the counterpart in real-world QCD.
  We predict the decay width of the techni-rho into weak bosons as $\Gamma_{\rho} / M_{\rho} \lesssim 0.24/N$,
  with the weak doublets number $N$; $N= 4$ (Farhi-Susskind model) or $N = 1$.
  We also show that the Weinberg spectral function sum rules
  and the Das-Mathur-Okubo (DMO) sum rule pole-dominated by the $\rho$ and $a_1$ are satisfied in our data,
  giving $S$ parameter and $L_{10}^r(M_\rho)$ consistent with the main result.
  Finally, we discuss 
  phenomenological implications for the BSM model building.
\end{abstract}

\maketitle


\section{Introduction}

The origin of mass is a central subject of particle physics today.
The Standard Model (SM) Higgs sector acquires a quadratic mass renormalization leading to the mass hierarchy problem,
which indicates the necessity of physics beyond the Standard Model (BSM).
Technicolor (TC) models---with mass being generated by spontaneous symmetry
breaking in a QCD-like theory---potentially solve the problem
but result in too-small flavor-changing neutral current (FCNC), too-heavy (unstable) Higgs bosons, and too-large \PTSprm
(see Refs~\cite{Matsuzaki:2015sya,Hill:2002ap} and references therein).
It is anticipated that all of these issues may be
overcome by assuming the technicolor interaction
to be approximately conformal, with a slow-running (``walking'') coupling and a large mass anomalous dimension ($\gamma_m$).
Such a theory, referred to as walking technicolor (WTC),
was originally proposed in a ladder Schwinger-Dyson (SD) approach
and resolved the FCNC problem~\cite{Yamawaki:1985zg} thanks to the large $\gamma_m$.
Moreover, a light Higgs boson naturally appeared as a dilaton~\cite{Yamawaki:1985zg,Bando:1986bg}
--- a pseudo Nambu-Goldstone (NG) boson associated
with the spontaneous breakdown of the scale symmetry\footnote{
Similar work has also been done without considering the anomalous dimension, scale symmetry,
or the dilaton that emerges from the breaking of the latter
~\cite{Holdom:1984sk,Akiba:1985rr,Appelquist:1986an}.}.

WTC may be realized by an asymptotically free QCD-like $SU(N_c)$
gauge theory with large number of massless flavors $N_f (\gg N_c)$~\cite{Appelquist:1996dq}, or ``large-$N_f$ QCD''.
In such theories,
the two-loop beta function has the Caswell-Banks-Zaks infrared fixed point (IRFP)~\cite{Caswell:1974gg,Banks:1981nn}
for large enough $N_f$, before losing asymptotic freedom.
It is very important to investigate the IR fixed point non-perturbatively.

Accordingly, 
in our previous work, we have performed
a series of studies of \SUN{3} lattice gauge theory with respect to its properties
at 
large $N_f$
~\cite{Aoki:2016wnc,Aoki:2014oha,Aoki:2013xza,Aoki:2013zsa,Aoki:2012eq,Aoki:2012ve,Aoki:2015zny,Aoki:2012yd}.
This work has been intended to explore the theory space from usual QCD
to theories in the conformal window, passing through the conformal phase boundary,
by selecting $N_f=4$, 8, 12, 16, in the fundamental representation.

Lattice studies have strongly suggested that a so-called conformal window emerges in asymptotically-free gauge theories
when the number of fermion flavors ($N_f$) exceeds a critical number $N_f^c$
(see Refs.~\cite{USQCD:2019hee,Witzel:2019jbe,Svetitsky:2017xqk,Pica:2017gcb} and references therein).
The conformal window is associated with the appearance of an IRFP
in the renormalization group flow~\cite{Caswell:1974gg,Banks:1981nn,Miransky:1996pd,Ryttov:2007sr}.
The gauge coupling is attracted to the IRFP and becomes independent of the renormalization scale.
In fact, such a property has been demonstrated in
large-$N_f$ QCD on the lattice
by using the running coupling defined by the step-scaling function (for recent developments, 
see Refs.~\cite{Witzel:2024bly,Hasenfratz:2024fad,Hasenfratz:2023wbr,Hasenfratz:2022qan,Hasenfratz:2020vta,Hasenfratz:2020ess,Hasenfratz:2019puu,Fodor:2018uih}
and references therein).
The conformality of many-flavor QCD is also suggested by the hadron masses (sourced by a small fermion mass $m_f$)
showing mass-deformed hyperscaling~\cite{Miransky:1984ef,DelDebbio:2010ze,DelDebbio:2010jy}
with a finite and universal $\gamma_m$ against varying $m_f$.

We have previously carried out comprehensive lattice studies of eight-flavor QCD and observed
incomplete conformality~\cite{Aoki:2016wnc}:
hadron mass spectra follow a conformal-like scaling with a large anomalous dimension $\gamma_m \sim 1$,
which is, however, non-universal, the pion ($\pi$) and flavor-singlet scalar meson ($\sigma$)
having different $\gamma_m, (\sim 0.4 - 0.5)$.
This property was contrasted to twelve-flavor QCD,
which has been observed to have a universal $\gamma_m$,
indicating that theory lies in the conformal window~\cite{Aoki:2012eq}.
The eight-flavor hadron spectra have also been described by
a polynomial fit ansatz motivated by chiral perturbation theory (ChPT).
The incomplete conformality would be regarded as a walking signal on the lattice.

More recently, we extended our study of eight-flavor QCD to the flavor-singlet spectrum, including the $\eta'$ meson, using updated lattice data~\cite{LatKMI:2025kti}.
We found a characteristic flavor-singlet spectrum with a light scalar
and an $\eta'$ that becomes relatively heavier as $N_f$ increases.

The above lattice findings indicates that
eight-flavor QCD is particularly interesting as a possible underlying theory of WTC.
First, the observed large $\gamma_m\sim 1$
allows an extended technicolor (ETC) model to realize
both light and heavy SM particles, while also giving FCNCs consistent with observations.
Second, $\sigma$ appears as a {\it stable} bound state as light as a pion,
in sharp contrast to real-world QCD.\@
The light $\sigma$ is interpreted as a pseudo-dilaton;
the spontaneous breaking of the chiral symmetry also breaks the scale symmetry explicitly,
giving rise to a nonperturbative trace anomaly, and to a small-finite $\sigma$-meson mass
in the chiral limit~\cite{Matsuzaki:2015sya}.
We have previously shown that the $\sigma$ meson could be responsible
for the discovered Higgs boson with 125 GeV mass~\cite{Aoki:2016wnc,Aoki:2014oha}.
The matter content of the $N_f = 8$ theory
matches the Farhi-Susskind (FS) model~\cite{Dimopoulos:1979sp,Farhi:1980xs},
which is the simplest ETC model with anomaly-free charge assignments;
while previously this model was rejected due to the problems with non-walking
technicolor models described above, the fact that eight-flavor QCD is walking
makes this model viable once again\footnote{
The chiral symmetry breaking \SUNLR{8}$\to$\SUNV{8}
gives rise to 63 NG-bosons (techni-pions)
60 of which have mass
given through ETC interactions by the  chiral condensate
enhanced with a large $\gamma_m$, thus not to be observed at present collider experiments,
while the remaining 3 are eaten by weak bosons.
See Ref.~\cite{Matsuzaki:2015sya}.
}
.

Moreover,
$SU(N_c)$ supersymmetric gauge theories with large $N_c$ have been investigated~\cite{Murayama:2021xfj,Murayama:2021rak}
to locate the edge of the conformal window.
Combined with anomaly mediated supersymmetry breaking,
non-supersymmetric gauge theories are non-perturbatively examined by taking advantage of Seiberg duality
and suggested to exhibit the spontaneous breaking of the chiral symmetry for $N_f\leq 3N_c - 1$~\cite{Murayama:2021rak}.
Assuming $N_c = 3$ is in the range of applicability,
$N_f = 8$ is the upper limit for the chiral symmetry breaking,
which supports the emergence of the walking feature at $N_f = 8$
indicated in our lattice studies~\cite{Aoki:2016wnc,Aoki:2014oha}.

In this paper, we extend these lattice studies
to the Peskin-Takeuchi $S$ parameter\footnote{
 Here we discuss the \Sprm per one weak doublet without subtraction of the SM contribution, unless otherwise stated.}
\cite{Peskin:1991sw} using the same $N_f=8$ ensembles as in our previous studies,
including the updated data used in our recent flavor-singlet spectrum study~\cite{LatKMI:2025kti}.\@ (Preliminary results were shown in Ref.~\cite{LatKMI:2016rjb}.)
The electroweak precision test~\cite{ParticleDataGroup:2020ssz} shows a tiny or vanishing $S$ parameter,
which has been one of the most important experimental
constraints on TC (and BSM strong dynamics more generally) applied to electroweak symmetry breaking;
if a naive scale-up of QCD was assumed as a BSM theory,
one would encounter too large an \Sprm due to the significant chiral symmetry breaking.
There is a prospect that the walking dynamics of eight-flavor QCD relaxes this problem.

The first lattice calculations of the \Sprm calculations used
overlap fermions \cite{Shintani:2008qe} and domain-wall fermions
\cite{Boyle:2009xi} on conventional QCD configurations.
To control chiral symmetry at finite lattice spacing
is essential for the lattice computation of the $S$ parameter,
which is the reason why the overlap and domain-wall fermions have been adopted.
In the context of strong dynamics theories to describe electroweak symmetry breaking,
the \Sprm has been studied by using ladder SD equation and Bethe-Salpeter (BS)
equation~\cite{Harada:2005ru,Kurachi:2006mu},
which indicates decreasing tendency of \Sprm when approaching the conformal window.

The Lattice Strong Dynamics (LSD) Collaboration
carried out the first systematic lattice study
of the $N_f$-dependence of $S$, using domain-wall fermions~\cite{Appelquist:2010xv,Appelquist:2014zsa}.
Their results indicate a decreasing trend of $S$ towards lighter fermion mass $m_f$ for $N_f=6$ and 8,
which matches the criteria described above for the BSM models built with these theories to be compatible with experiment.
The trend is also claimed to be backed up by the tendency of the observed spectrum towards parity doubling.
However, our previous work for $N_f = 8$ spectra~\cite{Aoki:2016wnc}
indicates an obvious deviation from the parity doubling,
and the later result of the LSD collaboration itself~\cite{LatticeStrongDynamics:2018hun}
finds consistent results with our results.
Our eight-flavor ensembles contain various volumes,
which allows advanced analyses on the finite volume (FV) effects on $S$.
We will show that both the $S$ parameter suppression and the parity doubling results from FV effects.

We make use of the exact chiral symmetry on two species of staggered fermions, which corresponds to eight flavors.
We measure the vector and axial-vector current correlators.
The \Sprm is extracted from the zero-momentum slope of the transverse vacuum polarization (VP, $\Pi(q^2)$),
i.e.\ the Fourier transform of the difference of the two correlators ($V-A$) by adopting two methods.
The first one is the traditional estimate: we fit $\Pi(q^2)$ against $q^2$ and
read off the slope $\propto S$ (VP-momentum method).
The second method utilizes the so-called time-momentum representation,
which has been developed in the precision science of the muon anomalous magnetic moment~\cite{Bernecker:2011gh}:
the \Sprm is directly evaluated in coordinate space by utilizing the time-moment of the $V-A$ correlators (TM method).
This work is the first to apply this TM method to the calculation of the $S$ parameter.
Comparisons of the two approaches give deeper understanding of the systematic uncertainties.
We examine the FV-corrected $S$ by developing a ChPT-motivated model, which allows us to extract reliable $S$ at small mass region.

Moreover, we evaluate vector/axial-vector meson decay constants and investigate
the Kawarabayashi-Suzuki-Riazuddin-Fayyazuddin
relations (KSRF-I, II), the Weinberg spectral function sum rules
(WSR-I, II), and the related Das-Mathur-Okubo (DMO) sum rule for the $S$ parameter.
In the presence of approximate conformality, whether the KSRF relations
and the Weinberg spectral function sum rules hold in eight-flavor QCD is an interesting and non-trivial question.

In the following section,
we explain our simulation setup and numerical measurement details.
Next in Sec.~\ref{sec:s_mom}, we show the \PTSprm calculated in the VP-momentum method.
In Sec.~\ref{sec:tmr}, we detail the TM method applied to $S$ and show the corresponding results.
In Sec.~\ref{sec:decay_const}, we investigate the vector/axial-vector meson decay constants and related quantities.
Then in Sec.~\ref{sec:discuss}, we discuss FV corrected $S$ and phenomenology.
Finally in Sec.~\ref{sec:summary}, we summarize our results and provide a future perspective.

\clearpage

\section{Vacuum Polarization Functions and $S$ Parameter on the Lattice }\label{sec:basics}

The \PTSprm is defined in the continuum theory
as the zero-momentum slope of the transverse vacuum polarization (VP, $\Pi(q^2)$),
i.e.~the Fourier transform of the difference between the vector and axial vector correlators,
\begin{eqnarray}
 S & = & 4\pi \Pi'(q^2=0), \label{eq:S}\\
 \Pi_{\mu\nu}(q) & = & \delta_{\mu\nu}\Pi(q^2) 
  -\frac{{q}_\mu{q}_\nu}{{q}^2}\tilde{\Pi}(q^2),\label{eq:Pi}\\
 \Pi_{\mu\nu}(q) & \equiv & \int d^4x e^{iqx}
  \left\{ \langle V^{(1,2)}_\mu(x) V^{(2,1)}_\nu(0)\rangle
   - \langle A^{(1,2)}_\mu(x) A^{(2,1)}_\nu(0)\rangle \right\}\ ,\label{eq:Pi_mn}
\end{eqnarray}
where $V^{(i,j)}_\mu$ and $A^{(i,j)}_\mu$ are the flavor non-singlet
vector and axial-vector current respectively, made of the anti-quark
field with $i$-flavor and the quark field with $j$-flavor.
This defines the contribution to the $S$ parameter of one electroweak doublet.

On the lattice, a power divergent contribution \cite{Shintani:2008qe} may arise in the
difference if the formulation breaks chiral symmetry explicitly.
One needs to use a formulation to compute the $S$ parameter on the lattice that prevents such an unphysical contamination.
In the literature, overlap fermions or domain wall fermions have been used
to formulate the $S$ parameter computation
on the lattice \cite{Shintani:2008qe,Boyle:2009xi,Appelquist:2010xv,Appelquist:2014zsa}.
Here we utilize the exact symmetry of the staggered fermion as a solution to this problem~\cite{Aoki:2016bfp}.

In order to fully utilize this symmetry to ensure that the divergence is exactly canceled
one needs to use the operators with
the following staggered taste-spin structures
\begin{equation}
 \begin{array}{cc} 
 (\gamma_\mu\otimes 1)\tau^a, & (\gamma_\mu\gamma_5\otimes\xi_5)\tau^a\\
 \end{array}
 \label{eq:structure}
\end{equation}
for the vector and axial vector respectively. 
See for example, Refs.~\cite{Gupta:1990mr,Gupta:1997nd} for the notation.
{
The eight-flavor system is realized as two staggered species, each providing four staggered tastes,
$SU(2)_{\rm species}\times SU(4)_{\rm taste}$,
and $\tau^a$ in Eq.~(\ref{eq:structure}) refers to the $SU(2)_{\rm species}$ generators
which bind a staggered antifermion field and a staggered fermion field.
}
The exact chiral symmetry with the generator
$(\gamma_5\otimes\xi_5)\tau^a$ transforms one into the other.
It is this transformation property that ensures the cancellation of the
UV power divergence when the $V-A$ difference of the vacuum polarization
functions is composed. 
The actual form of the operators depends on the choice of staggered action.

As in our previous work, we use the highly improved staggered quark (HISQ) action~\cite{Follana:2006rc,MILC:2010pul} with 
the tree-level Symanzik-improved gauge action~\cite{Symanzik:1983dc}.
The explicit form of the action and its properties are detailed in Ref.~\cite{Aoki:2016wnc}. 
The parameters used and statistical information about the ensembles are summarized in Table \ref{tab:meas_va}
and will also be made available digitally in Ref.~\cite{datapackage}.
Eight degenerate flavors are simulated via two species of staggered fermion; thus the fermionic determinant can be calculated without the need for a fourth root as is needed in theories with smaller numbers of non-degenerate flavors.
There is an exact $SU(2)_L\times SU(2)_R$ symmetry on the lattice,
which is expected to become full $SU(8)_L\times SU(8)_R$ in the continuum limit.
A convenient measure of the violation of the full chiral symmetry is the
taste symmetry breaking in the pion sector. The HISQ formulation is designed to reduce the taste symmetry breaking in simulations of real-life QCD. 
It appears also for our eight-flavor simulation that taste symmetry
is largely unbroken, with violations as measured by the squared pion mass being at most 5\%
\cite{Aoki:2016wnc}.
Note that the cancellation of the UV divergence in the $S$ parameter 
is unrelated to the taste symmetry, and is always guaranteed 
if one uses the proper set of operators.

In Appendix \ref{app:hisq_current},
we summarize the currents that we use to compute the vacuum polarization function for HISQ;
here we only describe the notation.
The ``conserved'' currents associated with the exact symmetry
in HISQ involve both length-one and -three point-split bilinears,
all having the spin-taste structure in Eq.~\eqref{eq:structure}. 
We use these currents denoted as $\Vc_\mu^{(i,j)}(x)$ and $\Ac_\mu^{(i,j)}(x)$ 
(see Eqs.~\eqref{eq:hisq_ccv} and \eqref{eq:hisq_cca}) at the position $x$ (sink).
At the position $0$ (source), simpler length-one currents
with the (finite) renormalization factor $Z_A$ (i.e. $Z_A V_\mu^{(i,j)}(0)$ and $Z_A A_\mu^{(i,j)}(0)$ are used. 
The vector and axial-vector currents mix via a transformation associated
with the exact chiral symmetry; as such the renormalization factors are equal ($Z_V=Z_A$).
The length one currents are defined in Eq.~\eqref{eq:hisq_olcs}.

The $V-A$ vacuum polarization function that matches the continuum one in Eq.~\eqref{eq:Pi_mn} is then
\begin{equation}
 \Pi_{\mu\nu}(q) = \frac{1}{4}\sum_{x}e^{iqx}
  \left\{
   Z_A \langle \Vc^{(1,2)}_\mu(x) V^{(2,1)}_\nu(0)\rangle
   - Z_A \langle \Ac^{(1,2)}_\mu(x) A^{(2,1)}_\nu(0)\rangle
  \right\}\ ,
  \label{eq:HISQ_VPF}
\end{equation}
where the factor of $1/4$ accounts for the two-species degrees of freedom.
The use of the conserved current at the sink position $(x)$ allows us to have
the Ward-Takahashi identities involving these correlation functions,
which are useful for checking our computation.
We keep the source operator as length one point-split, which is non-conserved,
to make the computational effort minimal and to have the coding as simple as possible.

The transverse vacuum polarization function $\Pi(q^2)$ is disentangled
using the continuum relation, Eq.~\eqref{eq:Pi},
once the vacuum polarization tensor $\Pi_{\mu\nu}(q)$ (Eq.~\eqref{eq:HISQ_VPF}) is computed on the lattice.

\begin{table}[!tbp]
\caption{
Parameters in the measurements of $V-A$ current correlators in eight-flavor QCD.
For each set of spatial ($L$), temporal ($T$) sizes and bare fermion mass in lattice units ($m_f$),
number of thermalized trajectories ($N_{\rm traj}$), trajectory interval for measurement ($N_{\rm traj}^{\rm int}$), number of configurations ($N_{\rm conf}$) analyzed, jackknife bin size ($N_{\rm traj}^{\rm bin}$) in trajectory unit, and number of measurement per configuration ($N_{\rm src}$) are summarized. Those with ``$\ast$'' indicate that the unit of trajectory is $\tau=0.5\times$ [molecular dynamics time unit], while the others are with $\tau=1$.
For this study we use a single Markov chain for each parameter set
selected from the multiple ones used in Ref.~\protect\cite{Aoki:2016wnc}.}
\label{tab:meas_va}
\begin{ruledtabular}
\begin{tabular}{llllllll}
\toprule
$L$ & $T$ & $am_f$ & $N_{\mathrm{traj}}$ & $N_{\mathrm{traj}}^{\mathrm{int}}$ & $N_{\mathrm{conf}}$ & $N_{\mathrm{traj}}^{\mathrm{bin}}$ & $N_{\mathrm{src}}$ \\
\midrule
48 & 64 & 0.009 & 8768* & 16* & 548* & 256* & 4 \\
42 & 56 & 0.012 & 4448 & 16 & 278 & 128 & 4 \\
42 & 56 & 0.015 & 2200 & 8 & 275 & 128 & 4 \\
36 & 48 & 0.015 & 7072 & 16 & 442 & 128 & 2 \\
36 & 48 & 0.02 & 10016 & 32 & 313 & 128 & 2 \\
30 & 40 & 0.02 & 16000 & 64 & 250 & 128 & 2 \\
30 & 40 & 0.03 & 10176 & 64 & 159 & 128 & 2 \\
24 & 32 & 0.03 & 47104 & 256 & 184 & 256 & 2 \\
30 & 40 & 0.04 & 14528 & 64 & 227 & 64 & 2 \\
24 & 32 & 0.04 & 70144 & 256 & 274 & 256 & 2 \\
18 & 24 & 0.04 & 2560 & 16 & 160 & 32 & 2 \\
18 & 24 & 0.06 & 2560 & 16 & 160 & 32 & 2 \\
18 & 24 & 0.08 & 2560 & 16 & 160 & 32 & 2 \\
18 & 24 & 0.1 & 2560 & 16 & 160 & 32 & 2 \\
\bottomrule
\end{tabular}

\end{ruledtabular}
\end{table}

The necessary renormalization factor for the length one operator is
calculated by taking the ratio:
\begin{equation}
 Z_A(t) = \frac{\langle {\mathcal A}_4(t) P(0)\rangle}
{\langle A_4(t) P(0)\rangle},
\end{equation}
where ${\mathcal A}_4(t)$ and $A_4(t)$ are the temporal component 
of the conserved and length one axial vector current respectively with
zero-momentum projection at time $t$. $P(0)$ is the pseudoscalar
operator with the structure $(\gamma_5\otimes\xi_5)$ at the origin. 
The axial current renormalization $Z_A$ for ${\mathcal A}_\mu=Z_A A_\mu$
is determined from the plateau at large $t$.

\begin{figure}[tbh]
\begin{center}
 \includegraphics[scale=0.94]{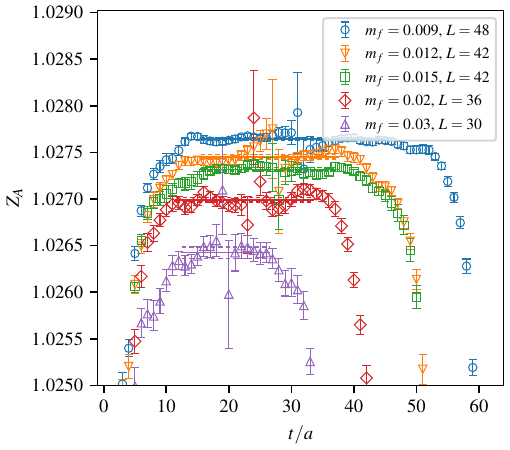}
 \hfill
 \includegraphics[scale=0.94]{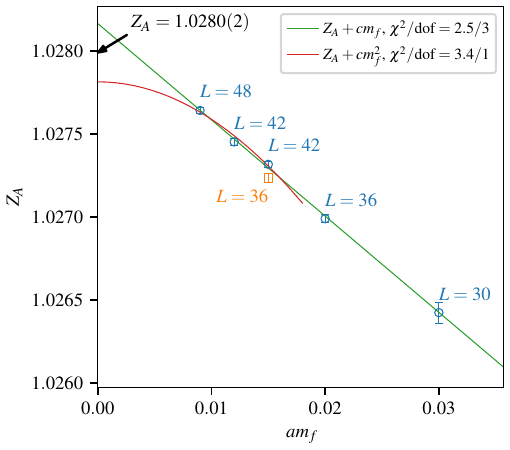}
\end{center}
\caption{Ratio and plateau fitting for $Z_A$ (left), and
  linear and quadratic chiral
 extrapolations in $m_f$ (right).}
\label{fig:ZA}
\end{figure}
The left panel of Figure \ref{fig:ZA} shows the ratio $Z_A(t)$ for the datasets with the smallest five masses.
For each mass, the largest volume has been adopted:
\begin{align}
(L,m_f) \in \ZAEnsembles\ ,
\end{align}
with $T = (4/3)L$.
The plateau emerges at large $t$ in each of the five data sets.
In the right panel, we plot the plateau values as a function of $m_f$.
The renormalization factor $Z_A$ is then determined by extrapolating to the chiral limit.
A linear extrapolation with all five points fits the data well.
To estimate the systematic error, we also consider an alternative 
extrapolation, $Z_A + c m_f^2$, using the three smallest masses.
As indicated in the right panel, we adopt the average of the two as our result
\begin{equation}
 Z_A = \ZAValue,
\end{equation}
where the error indicates the systematic one estimated from the difference
between this average and the linear or quadratic extrapolated value.
The size of the error is \ZASystematicPercentage.
The statistical error is as small as \ZAStatisticalPercentage, and thus is negligible.

\begin{figure}[tbh]
\begin{center}
\includegraphics[scale=0.94]{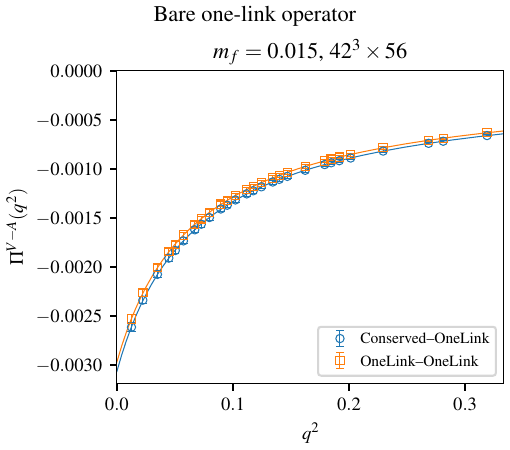}
\hfill
\includegraphics[scale=0.94]{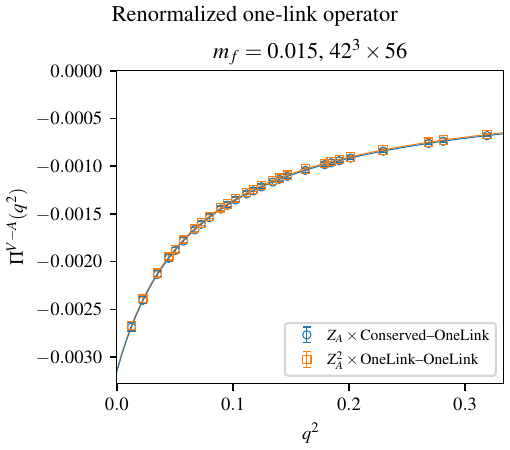}
\end{center}
\caption{The transverse vacuum polarization function as a function of momentum
 squared $q^2$ for $m_f=0.015$, $42^3\times 56$. The slight difference
 observed in the bare results (left) between those obtained with
 the conserved (blue circles) and the length one (one-link)
 current (orange squares) at the sink position disappears after all currents are
 renormalized (right), as expected. Note that these results are highly
 correlated.
}
\label{fig:test_ren}
\end{figure}

As a check whether the obtained renormalization factor $Z_A$ works as expected,
in Fig.~\ref{fig:test_ren}, we plot
the vacuum polarization for $(L,m_f) = (42, 0.015)$.
The data obtained with the conserved current
are slightly smaller than those from the length one current.
When multiplied by the renormalization factors,
the two become consistent.
This justifies the use of the non-conserved current at the source
position Eq.~(\ref{eq:HISQ_VPF}).
In later sections we only use the data with
the conserved current at the sink position (represented by blue circles in the figure)
to calculate the $S$ parameter.

\clearpage

\section{$S$ parameter from Pad\'e fit for $\Pi(q^2)$}\label{sec:s_mom}

For this section, let us assume
that the chiral symmetry breaking in eight-flavor QCD is the cause of the electroweak symmetry breaking observed in nature.
This symmetry breaking results in a finite $V-A$ current correlator, and thereby contributes to the Peskin-Takeuchi $S$ parameter.
We evaluate these effects for a single electroweak doublet.
The $S$ parameter as defined in Eq.~(\ref{eq:S}) requires the slope of the vacuum polarization $\Pi(q^2)$ at zero momentum,
which is not accessible directly from lattice data.
Therefore, we perform Pad\'e fits to obtain $\Pi(q^2)$, from which $S$ may be calculated~\cite{Appelquist:2010xv}.
In what follows, we refer to this approach as {\it VP-momentum method}.

\begin{figure}[tbh]
\begin{center}
 \includegraphics[scale=0.94]{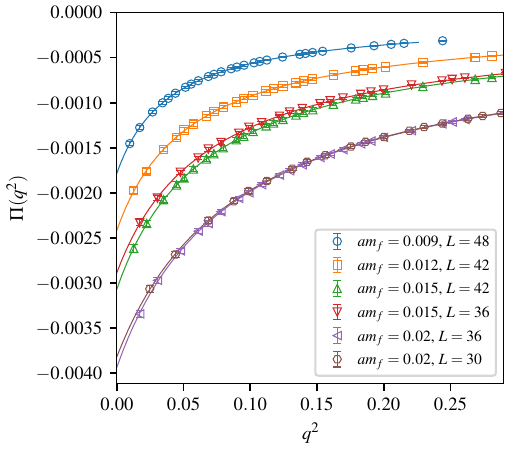}
 \hfill
 \includegraphics[scale=0.94]{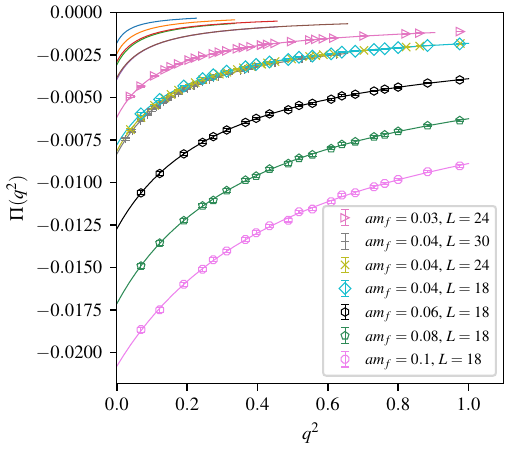}
\end{center}
\caption{The bare $V-A$ transverse vacuum polarization function $\Pi(q^2)$ (points)
 and the corresponding Pad\'e fit to Eq.~(\protect\ref{eq:pade}) (lines).
 Left: Results from lighter-mass ensembles ($m_f=\vpflightensemblesMassRange$).
 Right: Heavier-mass ensembles ($m_f=\vpfheavyensemblesMassRange$), including also the fits from the left panel.
} 
\label{fig:Pi-q2_fit}
\end{figure}

In Fig.~\ref{fig:Pi-q2_fit}, we show $\Pi(q^2)$ as a function of $q^2$.
These results are from the bare two-point functions for $V-A$.
The length one operator at the source position
must be renormalized; we multiply the obtained results by $Z_A$ to account for this.
In addition to the numerical results, we also show fits with the Pad\'e ansatz,
\begin{align}
f(q^2) = \frac{b_0 + b_1 q^2}{1 + c_1 q^2 + c_2 q^4}\ .\label{eq:pade}
\end{align}
To ensure the fit region is valid,
we constrain each momentum component to be
$|q_\mu|\le 2\times 2\pi/L_\mu$
(twice the unit momentum in each direction), and further limit the norm of the momentum to be $q^2<1$.
(The latter constraint only affects the smallest volume $18^3\times24$.) We then use the $\Pi(q^2)$ data for all
combinations of $q_\mu$ satisfying these constraints.
The fits give $\chi^2/\text{dof}\lesssim \PadeFitSecondLargestChisquare$ for all ensembles except the largest mass point $m_f=0.1$, where
$\chi^2/\text{dof} \simeq \PadeFitLargestChisquare$.
Here, the correlations among the data with different momenta
have not been considered.
This is sufficient for our purpose;
the results will be used for a consistency check
with a new approach (time-moment method) explained in the next section,
where the correlated fits will be adopted to extract our final results.
The systematic error associated with
the fit range is discussed in Appendix \ref{app:fit_range}.

\begin{figure}[tbh]
\begin{center}
 \includegraphics[scale=0.94]{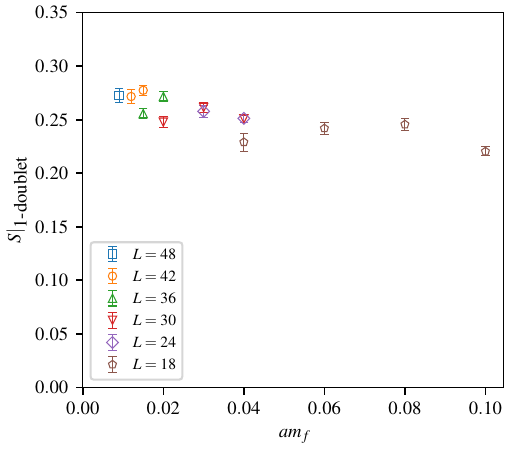}
 \caption{The strong-dynamics contribution to the $S$ parameter per
 electroweak doublet as a function of quark mass $m_f$.
 Colors and symbols indicate the lattice volume $L$ as shown in the legend.
 The error bars show the total uncertainty.}
\label{fig:S}
\end{center}
\end{figure}

The $S$ parameter calculated through the Pad\'e fit, and renormalized via multiplication by $Z_A$, is shown in Fig.~\ref{fig:S}
as a function of the quark mass $m_f$.
We find
\begin{align}
S = \SParameterVPZeroPointZeroZeroNine\qquad (m_f = 0.009)\ ,\label{eq:Smom0009}
\end{align}
and similar values for the largest volume at each quark mass $m_f \leq 0.02$.
In the heavier mass region,
the $S$ parameter tends to be suppressed:
the value of $S$ at the heaviest mass $m_f=0.1$ is over 20\% lower than that at the lightest mass $m_f=0.009$.
The detailed numbers are summarized in the fourth column of Table \ref{tab:s_lat}.

We may compare this result for the $S$ parameter with that which would be obtained by a
naive scale up of $N_f = 2$ QCD to the electroweak scale (without SM subtraction).
Equation~(5.78) of Ref.~\cite{Harada:2003jx},
and references therein for the axial vector form factor of $\pi\rightarrow e\nu \gamma$ and the pion charge radius,
gives an estimate of the low energy ChPT constant 
$\overline{L}_{10} = -0.0070(2)$. This in turn suggests that $S=0.35(1)$.
However, replacing the pion charge radius with the latest world average value~\cite{ParticleDataGroup:2024cfk},
we would instead obtain $\overline{L}_{10} = -0.0067(2)$, giving $S = 0.33(1)$.
To account for this uncertainty arising from the input pion charge radius, we adopt\footnote{
{
We also find a similar value 
$S=0.34(3)$ by using the renormalized low energy constant
$L_{10}^r(M_\rho) = -0.0055(7)$
reported in Ref.~\cite{Bijnens:1994qh},
which is based on the original ChPT estimates~\cite{Gasser:1983yg}
$\bar{l}_5 = 13.9(1.3)$ implying $\bar{L}_{10} = -0.0068(7)$.
}
}
\begin{align}
 S|_{N_f = 2} = 0.35(3)\ .\label{eq:SNf2}
\end{align}
Our $S$ parameters with $N_f = 8$ are smaller at all fermion masses.
More detailed discussions will follow in Sec.~\ref{GLparameter}.

Figure~\ref{fig:S} shows that the $S$ parameter measured at a particular lattice
volume bends down as the small-$m_f$ region is approached.
This movement is a finite volume (FV) effect, as simulating at the same mass on a
larger volume will move the point back upward.
In order to test these data against experiment, the standard
Higgs sector must be subtracted from this result~\cite{Peskin:1991sw,Schaich:2011qz}.
The FV effects observed here, however, are independent from this requirement.
We will discuss FV effects further in Sec.~\ref{sec:discuss}.

\clearpage

\section{$S$ parameter from time moments}\label{sec:tmr}

In this section, we investigate the Peskin-Takeuchi $S$ parameter
by utilizing the time moment of the $V-A$ current correlator.
A similar formulation has been developed in lattice QCD studies of the hadron vacuum polarization and its contribution to
the muon anomalous magnetic moment ($g_{\mu}-2$)~\cite{Bernecker:2011gh}.

\subsection{Time-Moment (TM) method}
We start from the scalar vacuum polarization $\Pi(q^2)$ defined in Eq.~(\ref{eq:Pi}).
In the rest frame, $q = (q_0,\vec{0})$, the scalar part is identical to
the spatially diagonal components of the tensor, $\Pi(q_0^2) = \Pi_{ii = 11,22,33}(q_0)$.
In this equality, the left-hand side is known to be an even function of $q_0$.
Therefore, the right-hand side $\Pi_{ii}(q_0)$ must be expressed as the cosine-type Fourier transformation
of the $V-A$ correlator. Thus, Eq.~(\ref{eq:Pi_mn}) reduces into
\begin{align}
\Pi(q_0^2) = \sum_{i = 1}^{3}\frac{\Pi_{ii}(q_0)}{3} = 2\int_0^{\infty}dt\ \cos(q_0 t)\ G^{\supvma}(t)
\ ,\label{eq:Pi_ft}
\end{align}
where $G^{\supvma}(t)$ denotes the $V-A$ current correlator obtained from
the right-hand side of Eq.~(\ref{eq:Pi_mn}) with zero-momentum projection,
\begin{align}
G^{\supvma}(t)
= \int d^3x \frac{1}{3}\sum_{i=1}^{3}
\bigl[
 \langle {V}^{(1,2)}_{i}(t,\vec{x})V^{(2,1)}_{i}(0)\rangle
 - \langle {A}^{(1,2)}_{i}(t,\vec{x})A^{(2,1)}_{i}(0) \rangle
\bigr]\ .\label{eq:VA}
\end{align}
A similar expression to Eq.~(\ref{eq:Pi_ft}) has been extracted in the context of muon $g-2$ physics,
and called the time-momentum representation (TMR)\footnote{
In the muon $g-2$ case, the hadron vacuum polarization function
from only the vector current correlator ($\Pi^V(q_0^2)$) is of interest.
The TMR is then applied to a dimension zero quantity,
$[\Pi^{V}(q_0^2)/q_0^2 - \Pi^{V}(k_0^2)/k_0^2]_{k_0\to 0}$.
The vector current conservation law guarantees that this is regular as $k_0\to 0$.
In the case of the $S$ parameter,
the TMR is instead defined for $\Pi^{\supvma}(q_0^2)$; as this has mass dimension $2$,
regularity is not a concern.
}
~\cite{Bernecker:2011gh,Francis:2013qkp,Francis:2013fzp}.

We shall now consider the TMR within lattice gauge theory.
The zero-momentum projected correlators read
\begin{align}
G^{\supvma}_{\rm lat}(t) &=
G^{{V}}_{\rm lat}(t) - G^{{A}}_{\rm lat}(t)\ ,\label{eq:VA_lat}\\
G^{{V}}_{\rm lat}(t) &=
Z_A\sum_{\vec{x}} \frac{1}{3}\sum_{i=1}^{3}
\langle {\mathcal{V}}^{(1,2)}_{i}(t,\vec{x}){V}^{(2,1)}_{i}(0)\rangle\ ,\label{eq:VV_lat}\\
G^{{A}}_{\rm lat}(t) &=
\sum_{\vec{x}} \frac{1}{3}\sum_{i=1}^{3}
Z_A\langle {\mathcal{A}}^{(1,2)}_{i}(t,\vec{x}){A}^{(2,1)}_{i}(0) \rangle\ ,\label{eq:AA_lat}
\end{align}
where $\mathcal{V}^{(1,2)}$ and $\mathcal{A}^{(1,2)}$ are the vector and axial vector currents,
defined more fully in Appendix~\ref{app:hisq_current}, and the suffix $(1,2)$
points to a pair of staggered species $(\bar{\phi}^{(1)},\phi^{(2)})$.
As the periodic boundaries of the lattice mean that correlators include both forward and backward contributions,
it is convenient to introduce a {\it folded} correlator defined in $t\in [0,T/2]$ as
\begin{align}
\hat{G}^{{X}}(t) =
\begin{cases}
G^{{X}}_{\rm lat}(t) & (t = 0\ \text{or}\ t = T/2)\ ,\\
\bigl(G^{{X}}_{\rm lat}(t) + G^{{X}}_{\rm lat}(T - t)\bigr)/2 & (0 < t < T/2)\ ,
\end{cases}
\label{eq:VA_fold}
\end{align}
where the suffix X is $V$, $A$, or $V-A$.
The lattice TMR reads
\begin{align}
\Pi_{\rm TMR}(q_0^2) = 2\sum_{t=0}^{T/2}~\cos(q_0 t)\ \hat{G}^{\supvma}(t)
\ ,\label{eq:Pi_ft_lat}
\end{align}
The time momentum $q_0$ is treated as a continuous variable rather than a discrete lattice momentum.
Therefore, $\Pi_{\rm TMR}(q_0^2)$ is not the strict Fourier transform of the correlator
in a finite box (and with a finite lattice spacing).
By specifying $q_0$ to be arbitrary small in Eq.~(\ref{eq:Pi_ft_lat}),
one can expand $\cos(q_0 t)$ around zero momentum
\begin{align}
\Pi_{\rm TMR}(q_0^2) =
2\sum_{t=0}^{T/2}\sum_{n=0}^{\infty}\frac{(-1)^n}{(2n)!}(q_0 t)^{2n}\hat{G}^{\supvma}(t)
\ .\label{eq:Pi_tmr}
\end{align}
As defined in Eq.~(\ref{eq:S}),
the $S$ parameter is evaluated from the derivative $[d\Pi(q_0^2)/dq_0^2]_{q_0\to 0}$,
which is extracted from the first ($n = 1$) time-moment contribution in Eq.~(\ref{eq:Pi_tmr}),
\begin{align}
S(T) = 8\pi\sum_{t=0}^{T/2}\frac{-1}{2}t^2\hat{G}^{\supvma}(t)\ .\label{eq:S_tmr}
\end{align}
It is this procedure that we will refer to as the Time-Moment (TM) method.

In this method, the $S$ parameter is directly calculated from the correlator $\hat{G}^{\supvma}(t)$,
neither working in momentum space, nor fitting the vacuum polarization function to obtain its zero-momentum slope.
However, there is a trade-off: we need to take the large $T$ limit in $S(T)$, and
model the $V-A$ correlator at larger distance $t > T/2$.

In Eq.~(\ref{eq:Pi_ft_lat}), $\Pi_{\rm TMR}(q_0\to 0) = 2\sum_{t = 0}^{T/2} \hat{G}^{\supvma}(t)$ gives a constant,
which acquires large finite volume (FV) effects.
Fortunately, the \Sprm in the TM method uses only the first moment in Eq.~(\ref{eq:Pi_tmr}),
where the FV effects coming from $q_0 = 0$ drop out. The remaining FV effect originates from the finite temporal extent $T$,
and the spatial zero-momentum projection. The former is addressed in the next subsection, while the latter is discussed
in Sec.~\ref{sec:decay_const}.

\subsection{Asymptotic behavior in Large-$T$}

In order to extract the $S$ parameter using the TM method, we must
develop a formulation to take the large $T$ limit in $S(T)$ given in Eq.~(\ref{eq:S_tmr}).
To this end, we decompose Eq.~(\ref{eq:S_tmr}) into UV and IR parts:
$\sum_{t=0}^{T/2}\to \sum_{t=0}^{t_0-1} + \sum_{t_0}^{T/2}$.
For the UV part ($t < t_0$), we will keep using the simulation results
used in the above.
For the IR part ($t \geq t_0$), we replace the lattice correlator $\hat{G}^{\supvma}(t)$
with a model correlator $G_{\text{model}}^{\supvma}$,
which is an analytic function of $T$, so that we can take the limit $T\to \infty$.
The explicit form of $G_{\text{model}}^{\supvma}$ will be specified in the later part of this section.
Here, we enumerate the procedure to calculate $S$:
\begin{enumerate}
\item
We require $G_{\text{model}}^{\supvma}$
to explain the lattice data in the region $t \in [t_0,t_1]$ via fit analyses.
The model becomes a function of the fit range, $G_{\text{model}}^{\supvma}(t, t_0, t_1, T)$.
\item
We construct the model $S$ parameter associated with $G_{\text{model}}^{\supvma}$
in the TM method,
\begin{align}
S_{\text{IR}}^{\text{model}}(t_0, t_1, \bar{t}, T) =
8\pi \sum_{t = t_0}^{\bar{t}} \frac{-1}{2!} t^2 G_{\text{model}}^{\supvma}(t, t_0, t_1, T)
\ .\label{eq:S_IR_model}
\end{align}
Note that the summation starts at the UV/IR threshold $t_0$ and ends up with $t = \bar{t}$,
which will ultimately be fixed to $T/2$, but is treated as a free parameter for the meantime.
\item
Now the $S$ parameter including both UV and IR contributions is rewritten as
\begin{align}
\bar{S}(t_0, t_1, \bar{t}, T)
= S_{\text{UV}}(t_0) + S_{\text{IR}}^{\text{model}}(t_0, t_1, \bar{t}, T)
\ ,\label{eq:S_approx}
\end{align}
where the first term is responsible for the contributions from the UV region $t\in [0,t_0)$
and evaluated with the folded lattice correlator Eq.~(\ref{eq:VA_fold}),
\begin{align}
S_{\text{UV}}(t_0) =
8\pi \sum_{t = 0}^{t_0 - 1} \frac{-1}{2!} t^2 \hat{G}^{\supvma}(t)
\ .\label{eq:S_UV}
\end{align}
\item
Then, we perform a large $T$ extrapolation by taking advantage of the analytic feature of the model,
and find the asymptotic $S$ parameter,
\begin{align}
\bar{S}(t_0, t_1) = \lim_{T\to\infty} \bar{S}(t_0, t_1, \bar{t} = T/2, T)\ .\label{eq:S_tmr_largeT}
\end{align}
\item
Finally, we average out $S(t_0,t_1)$ over the various choices of the fit ranges $[t_0, t_1]$,
\begin{align}
S = \langle \bar{S}(t_0,t_1)\rangle_{t_0,t_1}\ .\label{eq:S_fin0}
\end{align}
The associated fluctuation is taken to be the systematic uncertainty.
\end{enumerate}
For readability, we summarize all temporal variables:
\begin{align}
t &:~\text{Source-sink separation in the lattice correlator $\hat{G}^{\supvma}$}\nn\\
  &\qquad \text{or the counterpart in the model correlator $G_{\text{model}}^{\supvma}$,}\nn\\
t_0 &:~\text{UV/IR threshold, corresponding to the lower bound of the fit range for $G_{\text{model}}^{\supvma}$,}\nn\\
t_1 &:~\text{The upper bound of the fit range, satisfying $<T/2$,}\nn\\
\bar{t} &:~\text{The upper bound by which the model $G_{\text{model}}^{\supvma}$ will be summed
to get $S_{\text{IR}}^{\text{model}}$,}\nn\\
T &:~\text{Originally, the lattice temporal extension, but instead a free parameter in the model.}\nn
\end{align}

\begin{figure}[!th]
\begin{center}
\includegraphics{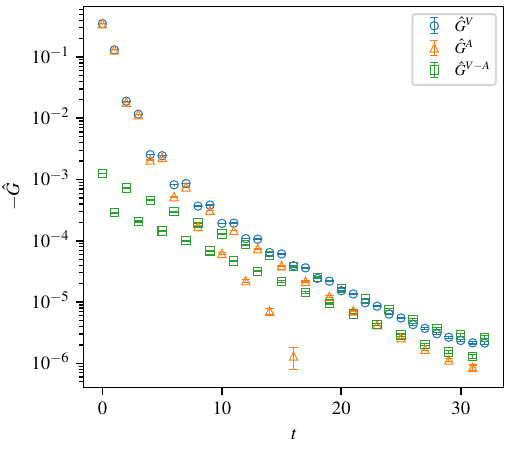}
\end{center}
\caption{
The vector and axial vector current correlators (blue circles and orange triangles respectively),
and their difference (green squares) for $m_f = 0.009$.
}\label{fig:corr_ens0}
\end{figure}

In order to find a suitable expression of the model correlator $G_{\text{model}}^{\supvma}$,
we shall examine the properties of our lattice correlators $\hat{G}^{{X = V,A,V-A}}(t)$
which have been defined in Eqs.~(\ref{eq:VA_lat}) -- (\ref{eq:VA_fold})
and shown in Fig.~\ref{fig:corr_ens0} in the $m_f = 0.009$ case.
The vector meson mass $M_{\rho}$ is smaller than the two-pion state
with the lowest momentum ($2\sqrt{M_{\pi}^2 + (2\pi/L)^2}$).
Therefore, the ground state in the vector correlator $\hat{G}^{{V}}(t)$
is always the lowest vector mode, which dominates at large distance.
The small oscillation seen in $\hat{G}^{{V}}(t)$ stems from the staggered parity partner.
In the axial vector correlator $\hat{G}^{{A}}(t)$,
the staggered partners are the vector states,
which include lighter modes than the lowest axial vector mode.
As a result, the staggered oscillation becomes dominant with increasing time separation $t$.
The above properties of the vector and axial vector currents hold in all ensembles.

We shall focus on the difference of the two correlators $\hat{G}^{\supvma}(t)$.
If we pick up data points at only even (or odd) $t$,
the slope is approximately constant across a wide range of $t$ in the logarithm plot.
The even-odd oscillation amplitude is also approximately constant.
This is interpreted as the large cancellation
of the excited modes between vector and axial vector,
which is advantageous for stable fits with wider and various ranges of $t$.
We shall consider the fit model for $\hat{G}^{\supvma}(t)$,
\begin{align}
G_{\text{model}}^{\supvma}(t, t_0, t_1, T)
&= (C_+^V(t_0,t_1) - (-1)^t C_+^A(t_0,t_1))
 \Bigl(\exp\bigl[-M_{\rho}t\bigr] + \exp\bigl[-M_{\rho}(T - t)\bigr]\Bigr)\nonumber\\
&- (C_-^A(t_0,t_1) - (-1)^t C_-^V(t_0,t_1))
 \Bigl(\exp\bigl[-M_{a_1}t\bigr] + \exp\bigl[-M_{a_1}(T - t)\bigr]\Bigr)\ .\label{eq:cva_large_t}
\end{align}
We fit the model $G_{\text{model}}^{\supvma}(t, t_0, t_1, T)$
to the lattice data $\hat{G}^{\supvma}(t)$ in the range of $[t_0, t_1]$,
where the correlations among data at different $t$ are taken into account.
The coefficients $\{ C_+^V, C_-^A, C_+^A, C_-^V\}(t_0, t_1)$ are the fit parameters
which show minor dependence on the fit range.
For the exponents $M_{\rho}$ and $M_{a_1}$ in the model,
we use the $\rho$-meson mass spectrum obtained in our previous work~\cite{Aoki:2016wnc},
and the $a_1$-meson mass spectrum from a concurrent,
orthogonal analysis~\cite{LatKMI:2025kti}, respectively\footnote{
In the next section, we investigate $M_{\rho}$ and $M_{a_1}$ using the correlators of the present work;
however, these show large statistical noise, and so are not included in the modelling of the $S$ parameter here.
}. These are summarized in Appendix~\ref{app:spectra_ref}.

Substituting Eq.~(\ref{eq:cva_large_t}) into (\ref{eq:S_IR_model}) and taking $T\to\infty$,
the asymptotic $S$ parameter Eq.~(\ref{eq:S_tmr_largeT}) reads
\begin{align}
&\bar{S}(t_0,t_1) =
 \bar{S}_{\text{UV}}(t_0)
 + \bar{S}_{\text{IR}}^{\text{V}+}(t_0, t_1)
 + \bar{S}_{\text{IR}}^{\text{A}-}(t_0, t_1)
 + \bar{S}_{\text{IR}}^{\text{A}+}(t_0, t_1)
 + \bar{S}_{\text{IR}}^{\text{V}-}(t_0, t_1)
\ ,\label{eq:S_fin}\\
&\Bigl(
\bar{S}_{\text{IR}}^{\text{V}+},
\bar{S}_{\text{IR}}^{\text{A}-},
\bar{S}_{\text{IR}}^{\text{A}+},
\bar{S}_{\text{IR}}^{\text{V}-}
\Bigr)(t_0, t_1)
=
4\pi\frac{-1}{2!} \Bigl(
C_+^V(t_0, t_1) I(t_0, M_{\rho}), -C_-^A(t_0, t_1) I(t_0, M_{a_1}), \nonumber \\
&\mathbin{\hphantom{=}}\hphantom{4\pi\frac{-1}{2!} \Bigl(}-C_+^A(t_0, t_1) J(t_0, M_{\rho}), C_-^V(t_0, t_1) J(t_0, M_{a_1})
\Bigr)\ ,\label{eq:S_each}
\end{align}
where the definitions
\begin{align}
I(t_0, M) &= \sum_{t = t_0}^{T\to\infty} t^2 e^{-Mt} = 
\frac{e^{-M(t_0 - 1)}}{(e^M - 1)^3} \bigl[t_0^2(e^M - 1)^2 + 2t_0(e^M - 1) + (e^M + 1)\bigr]\ ,\\
J(t_0, M) &= \sum_{t = t_0}^{T\to\infty} (-1)^t t^2 e^{-Mt} = 
\frac{(-1)^{t_0}e^{-M(t_0 - 1)}}{(e^M + 1)^3} \bigl[t_0^2(e^M + 1)^2 - 2t_0(e^M + 1) - (e^M - 1)\bigr]\ ,
\end{align}
originate from the forward correlators in our model function (\ref{eq:cva_large_t}).
The backward contributions proportional to $\exp[-M_{\rho}(T - t)]$ are suppressed as $T\to\infty$.
Once the fit coefficients $\{C_+^V, C_-^A, C_+^A, C_-^V\}(t_0,t_1)$ are determined by fits,
the above formulas allow us to evaluate the asymptotic $S$ parameter Eq.~(\ref{eq:S_fin0}).

\begin{figure}[!th]
\begin{center}
\includegraphics[scale=0.94]{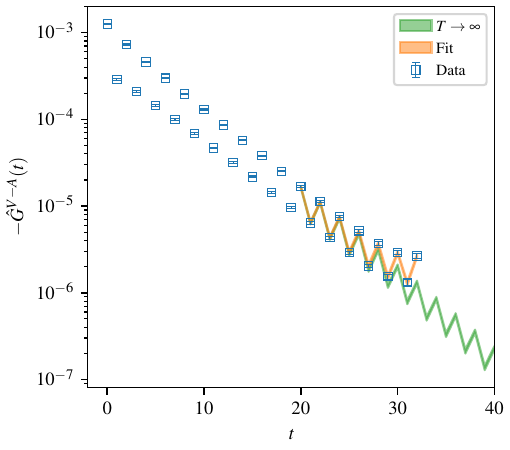}
\hfill
\includegraphics[scale=0.94]{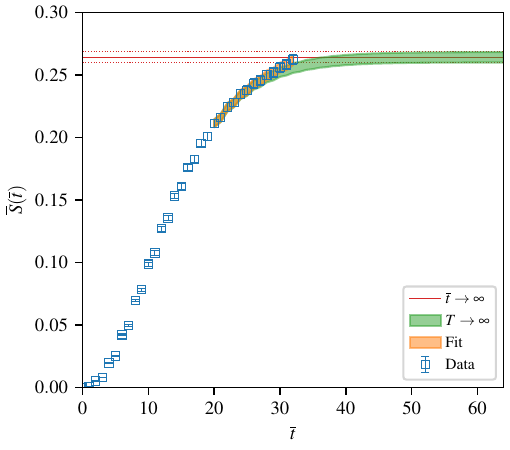}
\end{center}
\caption{
Left: The $V-A$ current correlator for $m_f = 0.009$.
The blue squares show the lattice data.
The green and orange lines represent
the fit function Eq.~(\protect\ref{eq:cva_large_t}) determined for the fixed fit range $[t_0,t_1] = \NfEightmfZeroPointZeroZeroNineTSixFourLFourEightPlateauRange$
with or without taking the large $T$ limit, respectively.
The width of the lines represents the statistical uncertainty.
Right: The accumulated $S$ parameter $\bar{S}(t_0,t_1,\bar{t}, T)$ defined in Eq.~(\protect\ref{eq:S_approx}).
The symbols are consistent with the left panel.
The blue squares are the lattice data obtained by replacing $T/2$ with $\bar{t}$ in Eq.~(\protect\ref{eq:S_tmr}).
The green and orange lines are obtained by substituting the relevant correlators from the left panel
into Eq.~(\protect\ref{eq:S_approx}).
The asymptotic value of the green line,
indicated by the red line,
corresponds to Eq.~(\protect\ref{eq:S_tmr_largeT}).
}\label{fig:fit_extrp}
\end{figure}

Figure~\ref{fig:fit_extrp} shows the fit result for $m_f = 0.009$.
In the left panel, the orange and green lines represent
the fit function Eq.~(\protect\ref{eq:cva_large_t}) determined in the range $[t_0,t_1] = \NfEightmfZeroPointZeroZeroNineTSixFourLFourEightPlateauRange$
with or without taking the large $T$ limit, respectively.
For a given $t > t_0$, the green line becomes slightly smaller than the orange one
since in the former the backward contributions $\propto \exp[-M_{\rho}(T - t)]$ vanish.
All fit parameters are well determined, with a good fit quality with $\chi^2/\text{dof} = \NfEightmfZeroPointZeroZeroNineTSixFourLFourEightChisquarePerDof$,
where the correlations among data at different $t$ have been considered.

Fits for various choices of $t_0 \geq \NfEightmfZeroPointZeroZeroNineTSixFourLFourEightPlateauStart$ and $t_1 \leq T/2$ work with $\chi^2/\text{dof} \sim \mathcal{O}(1)$,
and the corresponding fluctuation in the fit parameters is an order of magnitude smaller than the statistical error.
We have observed similar properties in the other ensembles, with suitable variations of the fit range $[t_0,t_1]$.
There is one exceptional ensemble: $m_f = 0.08$.
As will be explained later, the systematic error at this point might be underestimated.

The right panel of Fig.~\ref{fig:fit_extrp} also shows
the accumulated $S$ parameter defined by Eq.~(\ref{eq:S_approx}),
as a function of the upper bound in this accumulation, $\bar{t}$, as introduced in Eq.~(\ref{eq:S_IR_model}).
Both lattice data (replacing $T/2$ by $\bar{t}$ in Eq.~(\ref{eq:S_tmr}))
and fit curves (from substitution of the fitted $G_{\text{model}}^{\supvma}(t, t_0, t_1, T)$,
with and without the infinite $T$ extrapolation
--- orange and green lines, respectively, with a color scheme consistent to the left panel) are shown.

We shall focus on the extrapolated curve, whose asymptotic value at $\bar{t}\to\infty$
corresponds to Eqs.~(\ref{eq:S_tmr_largeT}) and (\ref{eq:S_fin}) with a specific fit range;
in the $m_f=0.009$ case this is $[t_0,t_1] = [20,32]$.
Taking account of the variation of the fit range, we obtain the final estimate of the \Sprm Eq.~(\ref{eq:S_fin0})
for this ensemble as $S = \NfEightmfZeroPointZeroZeroNineTSixFourLFourEightSParameter$. The first parenthesis shows the statistical uncertainty, while
the second represents the systematic error associated with the choice of fit range $[t_0,t_1]$.
The results for all ensembles are summarized in Table~\ref{tab:s_lat}.

\subsection{Results for \Sprm in TM method}

\begin{figure}[!th]
\begin{center}
\includegraphics{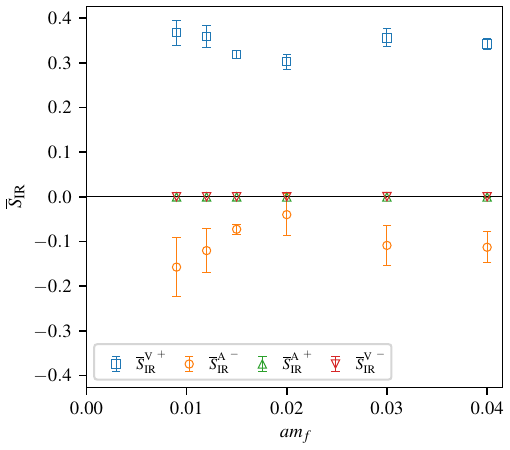}
\end{center}
\caption{
  The IR contributions to the $S$ parameter in the large $T$ limit,
  as defined in Eq.~(\protect\ref{eq:S_each}).
  The average over various fit ranges has already been taken.
  The error bars show the total uncertainties where the statistical uncertainties dominate.
}\label{fig:sparam_va_sys0}
\end{figure}

\begin{figure}[!th]
\begin{center}
\includegraphics{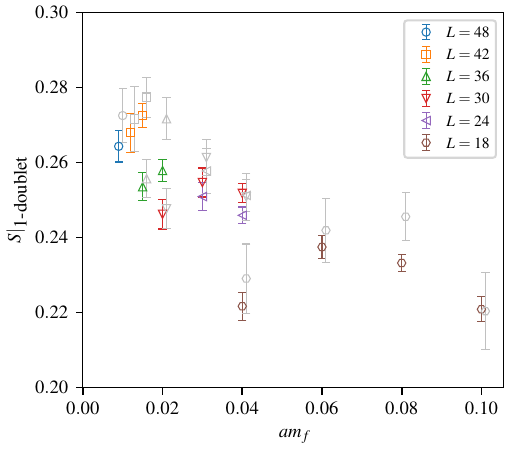}
\end{center}
\caption{
The $S$ parameter using the TM method (colored) as a function of $m_f$,
compared with the results of the VP-momentum method (gray).
The error-bars show the total uncertainty.
See text for the tension found at  $(L,m_f) = (36,0.02)$ and $(18,0.08)$.
}\label{fig:sparam_tmr_vs_mom}
\end{figure}

In Fig.~\ref{fig:sparam_va_sys0}, we show the partial IR contributions to the $S$ parameter defined in Eq.~(\ref{eq:S_each})
with the average over various fit ranges, for several selected ensembles.
The leading contribution is due to the ground state in the vector current correlator $\bar{S}_{\text{IR}}^{\text{V}+}$,
for which the statistical uncertainty dominates in all selected ensembles.
The next-to-leading contribution is from the ground state
in the axial vector current correlator $\bar{S}_{\text{IR}}^{\text{A}-}$, carrying a negative sign.
The remaining contributions stem from the oscillating parts
of the correlator and are practically negligible.
The $m_f$ dependence is mild in all cases.

In Fig.~\ref{fig:sparam_tmr_vs_mom},
we show the $S$ parameter in the large $T$ limit defined in Eq.~(\ref{eq:S_fin0}) for all of our ensembles.
The results obtained from the TM method, which we call $S_{\rm TM}$, are compared with
those obtained via the VP-momentum method $S_{\text{VP-Mom}}$ as discussed in the previous section.
In the former case, the uncertainty estimate includes statistical and
systematic uncertainties added in quadrature, where the systematic uncertainties
are estimated from the $T\rightarrow\infty$ treatment as described above.
{
The corresponding data are tabulated in Table \ref{tab:s_lat}:
$S_{\rm TM}$ (the right-most column) are consistent with $S_{\text{VP-Mom}}$ (the fourth column)
within the statistical errors in most cases
though $S_{\rm TM}$ tends to be slightly smaller.
There are two exceptions, $(L,m_f) = (36,0.02)$ and $(18,0.08)$,
where the two methods show the tension beyond the total error.}
In the former case, the tension significantly reduces
after taking account of the finite volume correction as will be discussed in the next section.
For the tension in the latter ensemble, there is a caveat:
the total error in $S_{\rm TM}$ is significantly smaller than that of $S_{\text{VP-Mom}}$.
For this ensemble the fit model Eq.~(\ref{eq:cva_large_t}) barely explains the lattice $V-A$ correlator for only three choices of the fit ranges.
This might be insufficient to estimate the associated systematics.
In practice, the region of interest is at small $m_f$, where this ensemble
will have a negligible contribution. As no other ensemble shows as poor a fit of the
model to the correlator, we choose not to pursue this issue further.

\begin{table}[!tbp]
\caption{
Summary of the electroweak one-doublet contribution to the $S$ parameter
obtained via the VP-momentum ($S_{\text{VP-Mom}}$) and TM ($S_{\rm TM}$) methods.
The first and second  parentheses denote the statistical
and systematic uncertainties, respectively.
The systematic uncertainty comes from the fit range variations
in the Pad\'e ansatz \protect\eqref{eq:pade} for $S_{\text{VP-Mom}}$
and the model function \protect\eqref{eq:cva_large_t} for $S_{\rm TM}$.
}\label{tab:s_lat}
\begin{ruledtabular}
\begin{tabular}{lllll}
\toprule
$L$ & $T$ & $am_f$ & $S_{\textrm{\scriptsize{VP-Mom}}}$ & $S_{\textrm{\scriptsize{TM}}}$ \\
\midrule
48 & 64 & 0.009 & 0.273(7)(2) & 0.264(4)(1) \\
42 & 56 & 0.012 & 0.272(7)(6) & 0.268(5)(1) \\
42 & 56 & 0.015 & 0.277(5)(2) & 0.273(3)(1) \\
36 & 48 & 0.015 & 0.256(5)(1) & 0.254(4)(1) \\
36 & 48 & 0.02 & 0.272(5)(3) & 0.258(3)(1) \\
30 & 40 & 0.02 & 0.248(5)(2) & 0.246(4)(0) \\
30 & 40 & 0.03 & 0.261(5)(2) & 0.255(4)(0) \\
24 & 32 & 0.03 & 0.258(5)(2) & 0.251(4)(0) \\
30 & 40 & 0.04 & 0.251(3)(5) & 0.252(3)(0) \\
24 & 32 & 0.04 & 0.251(4)(2) & 0.246(2)(0) \\
18 & 24 & 0.04 & 0.229(8)(4) & 0.222(4)(1) \\
18 & 24 & 0.06 & 0.242(6)(6) & 0.237(3)(0) \\
18 & 24 & 0.08 & 0.246(6)(2) & 0.233(2)(0) \\
18 & 24 & 0.1 & 0.220(4)(9) & 0.221(3)(0) \\
\bottomrule
\end{tabular}

\end{ruledtabular}
\end{table}

\clearpage

\section{Vector and axial-vector decay constants}\label{sec:decay_const}

In this section, we investigate the vector and axial-vector decay constants ($F_{\rho,a_1}$)
extracted from the same correlators $\hat{G}^{{V,A}}$ as used in the $S$ parameter studies.
We test the Kawarabayashi-Suzuki-Riazuddin-Fayyazuddin (KSRF) relations~\cite{Kawarabayashi:1966kd,Riazuddin:1966sw}
and Weinberg spectral function sum rules (WSR)~\cite{Weinberg:1967kj} against these data. 
We also study the $S$ parameter
through the Das-Mathur-Okubo spectral function sum rule (DMOSR)~\cite{Das:1967ek}.
Based on the results, we discuss the implications for the presence or otherwise of
chiral symmetry breaking in eight-flavor QCD.
We shed light on potential BSM phenomenology via the decay width of the techni-rho to weak bosons
mediated by two techni-pions.

\subsection{Setup}\label{subsec:decay_const_setup}

We shall express the correlator defined in Eqs.~(\ref{eq:VV_lat}) and (\ref{eq:AA_lat})
in the spectral representation,
\begin{align}
G_{\rm lat}^{{X=V,A}}(t)
 &= \frac{|Z_X|^2}{2M_0^{(X)}}e^{-M_0^{(X)}t} + \cdots\ ,\label{eq:FinG}\\
|Z_X|^2 &= \frac{1}{3}\sum_{i=1}^3
\langle 0 | X_i^{(1,2)}(0) | E_0^{(X)}(\vec{0}) \rangle
\langle E_0^{(X)}(\vec{0}) | X_i^{(1,2),\dagger}(0) | 0 \rangle
\ ,\label{eq:ZX}\\
X_{i}^{(1,2)}(x) &=
V_{i}^{(1,2)}(x)\ \text{or}\ A_{i}^{(1,2)}(x)\ ,\quad (i = 1,2,3)\ ,
\end{align}
where $|E_0^{(X)}(\vec{p})\rangle$ denotes
the ground state in the $X$ channel with 
momentum $\vec{p}$ and energy $E_0^{(X)}(\vec{p})$,
for which the mass is defined as $M_0^{{(X)}} = E_0^{(X)}(\vec{p} = \vec{0})$.
The ellipsis represents contributions from the excited one-particle, 
and multi-particle states.
The decay constants for one-doublet contributions are defined through
the overlap of the vector (axial-vector) ground state with the vacuum,
\begin{align}
F^2_{{X}} &=
\frac{1}{4}\biggl(\frac{|Z_X|}{M_0^{(X)}}\biggr)^2\ .\label{eq:FX}
\end{align}
The prefactor $1/4$ accounts for the staggered taste degeneracy.
Since we have measured one-species (i.e. 4 tastes) correlator,
the matrix element $\langle 0 | X_i^{(1,2)}(0) | E_0^{(X)}(\vec{0}) \rangle$ contains 4 taste contributions.
To extract one-doublet decay constant, we must replace
$\langle 0 | X_i^{(1,2)}(0) | E_0^{(X)}(\vec{0}) \rangle$
with $\langle 0 | X_i^{(1,2)}(0) | E_0^{(X)}(\vec{0}) \rangle/\sqrt{4}$~\cite{Kilcup:1986dg},
which results in the prefactor.

We extract the (axial)vector decay constants $F_X$ as well as masses $M_0^{(X)}$
by fitting the correlator data to Eq.~(\ref{eq:FinG}) and using Eq.~(\ref{eq:FX}).
As the decay constants are found by considering the vector and axial-vector currents separately,
unlike in the previous section, we cannot utilize the cancellations of the excited states in the
$V-A$ current correlators to improve the signal.
Then, 
the available variations of possible fit range become restricted.
The systematic error in $F_X$ and $M_0^{(X)}$ will be estimated by the fluctuation of fit parameters
associated with the fit range variations.

In the following, we will concentrate on the ensembles:
\begin{align}
(L, m_f) = \LightEnsemblesWithSystematics,\label{eq:lv_ens}
\end{align}
for which the fit range can be varied enough to estimate the associated systematic uncertainty.
This corresponds to the largest volume data for a given mass, with the exception of $m_f = 0.015$,
for which the $L = 36$ data is considered in preference to the larger $L = 42$ due to the
significantly larger statistics available.

We note that this choice of ensembles is consistent
with the one adopted in our previous precise study of the $N_f=8$ mass spectrum~\cite{Aoki:2016wnc},
with the exception of the lightest data $m_f = 0.009$, which is a new data point generated
for this work and the study of the flavor-singlet scalar mass~\cite{LatKMI:2025kti}.

\subsection{Results for $M_{\rho,a_1}$ and $F_{\rho,a_1}$}\label{subsec:res_f_ra}

\begin{figure}[tbh]
\begin{center}
 \includegraphics[scale=0.94]{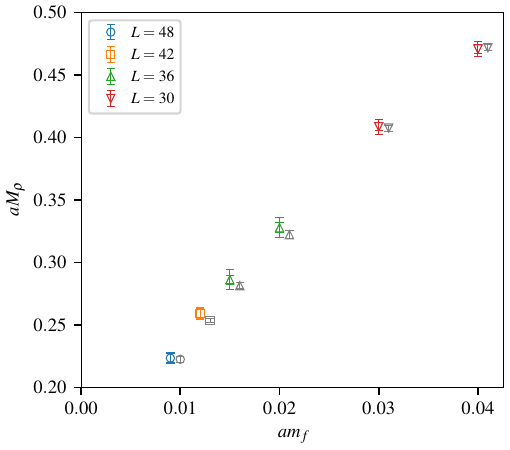}\hfill
 \includegraphics[scale=0.94]{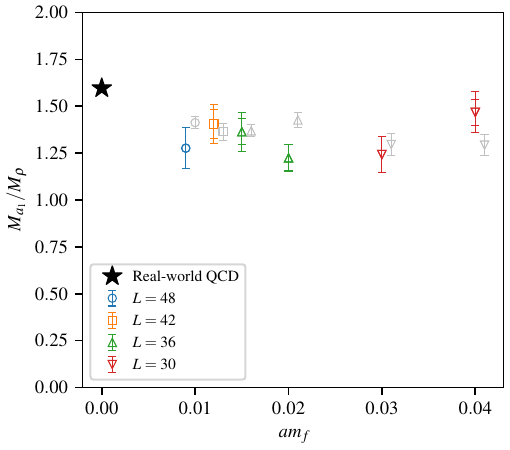}
\end{center}
 \caption{
 Left: Vector meson masses $M_\rho$ extracted from the length one
 point-split currents (colored) and the point operators (gray, from~\protect\cite{Aoki:2016wnc})
 plotted against quark mass. The latter points have been shifted horizontally for ease
 of comparison.
 For the colored symbols, inner and outer error bars
 represent the statistical and combined statistical and systematic uncertainties respectively.
 For the gray symbols, the error bars show the total uncertainties.
 Right: The ratio $M_{a_1}/M_\rho$ similarly to the left panel.
 The value for real-world QCD is also shown as a reference.
 }
\label{fig:vectormass}
\end{figure}

In Fig.~\ref{fig:vectormass}, we show the vector and axial-vector masses
($M_{\rho, a_1} \equiv M_0^{( X=V,A)}$) measured
with the length one point-split current operators in this work,
including a comparison with our previous results~\cite{Aoki:2016wnc,LatKMI:2025kti} from the PV-channel
measurement with point operators.
The difference of operators provides a measure of the staggered taste symmetry breaking.
We observe a tiny taste symmetry breaking, much smaller than the total errors.
Even in the pion sector,
where the spontaneous breaking of the chiral symmetry makes the Goldstone pion distinctive to
its taste partners, we have previously confirmed that the taste symmetry breaking
is tiny~\cite{Aoki:2016wnc}.
Therefore, good taste symmetry for the vector channel is a plausible observation.
The $\rho$ and $a_1$ mass spectra are more precisely determined in our previous study~\cite{Aoki:2016wnc,LatKMI:2025kti}.
Therefore, we have used them in the $S$ parameter computation via Eq.~(\ref{eq:cva_large_t}).
The small taste violations that we now observe justify this choice of strategy.

Also plotted in Fig.~\ref{fig:vectormass} is the ratio $M_{a_1}/M_\rho$ as a function of $m_f$.
The ratios (colored symbols)
are roughly consistent with our previous results measured
with different taste operators (gray symbols)~\cite{Aoki:2016wnc,LatKMI:2025kti}.
The previous results with smaller uncertainty allowed to
take the chiral limit and found $M_{a_1}/M_\rho\simeq 1.43\simeq \sqrt{2}$~\cite{Aoki:2016wnc}.
This was in sharp contrast to the ``parity doubling''
signal ($M_{a_1}/M_\rho \to 1$)~\cite{Appelquist:2010xv,Appelquist:2014zsa}
and indicated the spontaneous breaking of chiral symmetry in eight-flavor QCD\footnote{
We observed a similar behavior of $M_{a_1}/M_\rho$ also
for $N_f=12$, which is in the conformal phase. However, this is for a different reason;
see Appendix G and Section IV.E of Ref.~\cite{Aoki:2016wnc}.}.
The present result supports that observation;
as such, we interpret the symmetry breaking signal as physical, and independent of the taste violation artifacts.

\begin{figure}[tbh]
\begin{center}
 \includegraphics[scale=0.94]{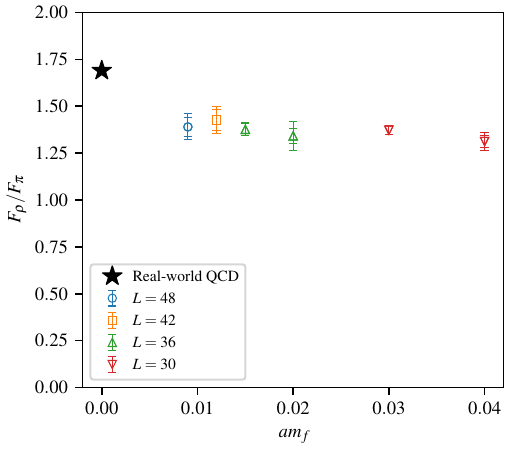}
 \includegraphics[scale=0.94]{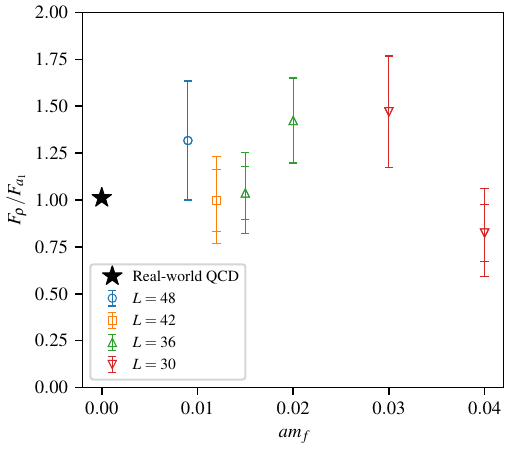}
\end{center}
\caption{
 Left: The vector-meson decay constant normalized by the pion decay constant.
 Right: The ratio of the vector-meson decay constant to the axial-vector meson decay constant.
 Inner and outer error bars
 represent the statistical and combined statistical and systematic uncertainties respectively.
}\label{fig:frho}
\end{figure}

In the left panel of Fig.~\ref{fig:frho}, we show the vector meson decay constant ($F_{\rho}$)
normalized by the pion decay constant $F_\pi$ ~\cite{Aoki:2016wnc,LatKMI:2025kti} (c.f.\ Appendix~\ref{app:spectra_ref}).
The ratio $F_{\rho}/F_{\pi}$ is approximately constant against $m_f$
and consistent with $\sqrt{2}$.
The ratio is also consistent with other work~\cite{LatticeStrongDynamics:2018hun}.
The right panel of Fig.~\ref{fig:frho} displays
the ratio $F_\rho / F_{a_1}$, which fluctuates against $m_f$ but remains consistent with $\sqrt{2}$
within the uncertainty.
These behavior can be explained by the KSRF relations and/or WSRs as detailed in the followings.
In Table~\ref{tab:va_spectra}, 
we summarize the numerical values for $M_{\rho,a_1}$ and $F_{\rho,a_1}$.

\begin{table}[!tbp]
\caption{
Summary table of vector and axial-vector masses and decay constants.
The first and second parentheses show the statistical and systematic uncertainties, respectively.
}\label{tab:va_spectra}
\begin{ruledtabular}
\begin{tabular}{rrlllll}
\toprule
$L$ & $T$ & $am_f$ & $F_{\rho}$ & $F_{a_1}$ & $M_{\rho}$ & $M_{a_1}$ \\
\midrule
48 & 64 & 0.009 & 0.0552(20)(19) & 0.042(11)(3) & 0.2235(38)(27) & 0.285(25)(5) \\
42 & 56 & 0.012 & 0.0648(26)(22) & 0.065(11)(13) & 0.2592(40)(30) & 0.364(20)(21) \\
36 & 48 & 0.015 & 0.0695(16)(5) & 0.067(10)(11) & 0.2862(35)(9) & 0.391(21)(23) \\
36 & 48 & 0.02 & 0.0785(23)(39) & 0.055(8)(0) & 0.3278(41)(58) & 0.402(23)(2) \\
30 & 40 & 0.03 & 0.0982(18)(1) & 0.067(14)(1) & 0.4087(31)(1) & 0.508(39)(2) \\
30 & 40 & 0.04 & 0.1084(26)(33) & 0.132(24)(32) & 0.4708(36)(39) & 0.691(32)(45) \\
\bottomrule
\end{tabular}

\end{ruledtabular}
\end{table}

\subsection{KSRF relation}\label{subsec:ksrf}

We examine the KSRF relations\footnote{
We note that our convention for $F_{\pi,\rho,a_1}$ contains a factor $\sqrt{2}$.
Substituting $F_{\pi,\rho,a_1} \to \sqrt{2}F_{\pi,\rho,a_1}$ gives
the form of the KSRF relations and WSRs used, for example,
in~Refs\protect\cite{Appelquist:2014zsa,LatticeStrongDynamics:2018hun}.},
\begin{align}
\text{KSRF-I:}\quad g_{\rho\pi\pi} &= \frac{M_{\rho}F_{\rho}}{\sqrt{2} F_{\pi}^2}\ ,\label{eq:ksrf1}\\
\text{KSRF-II:}\quad g_{\rho\pi\pi} &= \frac{M_{\rho}}{F_{\pi}}\ ,\label{eq:ksrf2}
\end{align}
where $g_{\rho\pi\pi}$ denotes the coupling between rho and two pions.
In the technicolor interpretation, the $\rho$ and $\pi$ mesons
are referred to as the technirho and technipions, respectively.

While KSRF-II is derived for a particular parameter choice
of Hidden Local Symmetry (HLS)~\cite{Bando:1984ej,Bando:1987br},
the KSRF-I in the chiral limit is a manifestation of the low energy theorem of the HLS~\cite{Bando:1985rf},
which has been
proven to all orders in nonlinear sigma models on an arbitrary compact group~\cite{Harada:1993jk, Harada:1993qi}.
Since our previous work~\cite{Aoki:2016wnc} indicates the chiral symmetry is spontaneously broken for $N_f = 8$,
the low energy theorem would be applicable.
As such, we assume that the KSRF-I holds in $N_f = 8$ QCD,
as it does in other theories with $N_f < 8$ that are in the broken phase.

\begin{figure}[tbh]
\begin{center}
\includegraphics[scale=0.94]{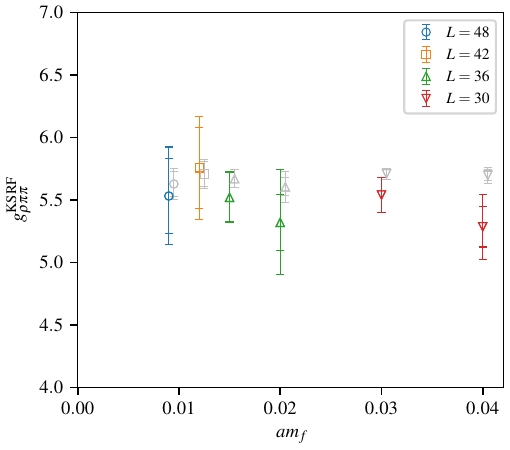}
\hfill
\includegraphics[scale=0.94]{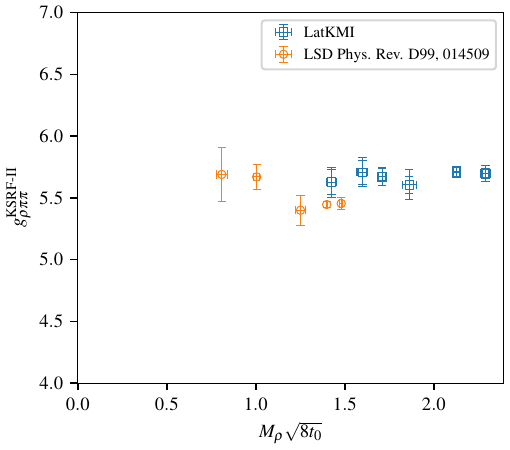}
\end{center}
 \caption{
 The coupling $g_{\rho\pi\pi}$ from the KSRF relations.
 Left: The colored and gray symbols are from KSRF-I (\protect\ref{eq:ksrf1}) and KSRF-II (\protect\ref{eq:ksrf2}) respectively;
 the latter are offset slightly to the right for ease of reading.
 Right: Our $g_{\rho\pi\pi}$ (blue squares) from KSRF-II compared with those given by the LSD collaboration
(\protect\cite{LatticeStrongDynamics:2018hun}, orange circles).
 The former is the same as the gray symbols in the left panel.
}\label{fig:ksrf}
\end{figure}

In Fig.~\ref{fig:ksrf} and Table~\ref{tab:ksrf_wsr}, we show $g_{\rho\pi\pi}$ from the two KSRF relations.
In the left panel, the results from KSRF-I (colored symbols) are independent of $m_f$ within uncertainties,
and imply that a similar value would arise in the chiral limit.
Unlike the KSRF-I, there is no theoretical justification
for the KSRF-II to hold in the present $N_f = 8$ QCD.
Therefore, it is interesting to ask whether the 
$g_{\rho\pi\pi}$ predicted from the KSRF-I
is reproduced by the KSRF-II.
In fact, the KSRF-I predictions (colored symbols)
become near-identical to the KSRF-II counterparts (gray symbols)\footnote{
The KSRF-I results in much larger errors;
this is due to very strong correlations between $F_\rho$ and $M_\rho$.}
at small $m_f$.
Thus, the KSRF-II is also satisfied, which is a non-trivial observation.
When both KSRF-I and II are satisfied as our data indicate,
the relation $a \equiv (F_\rho/F_\pi)^2 = 2$ results
from equating the right-hand side of two KSRFs (Eqs.~\eqref{eq:ksrf1} and \eqref{eq:ksrf2}).
We note that a theoretical proof for ``$a = 2$'' has been shown only for $N_f = 2$
in the large $N$ limit of the Grassmannian $O(N)/[O(N-3)\times O(3)]$
nonlinear sigma model~\cite{Yamawaki:2018jvy,Yamawaki:2023ybn} using HLS~\cite{Bando:1984ej,Bando:1987br}.
In the present system, ``$a$'' remains as a parameter while we have numerically found $a\simeq 2$
as already shown in the left panel of Fig.~\ref{fig:frho}.
Thus, the KSRF-II is found to hold even for $N_f = 8 \neq 2$, which is non-trivial.

In the right panel of Fig.~\ref{fig:ksrf},
we compare our $g_{\rho\pi\pi}$ (blue squares) from the KSRF-II
with those given by the LSD collaboration (orange circles)~\cite{LatticeStrongDynamics:2018hun,Appelquist:2016viq}.
The horizontal axis shows $\rho$ meson masses normalized by the Wilson flow scale $t_0$.
For our data, we used $M_{\rho(PV)}$ shown in Tables~\ref{tab:spectra_ref} in Appendix~\ref{app:spectra_ref}
and $t_0$ in our work~\cite{LatKMI:2025kti}.
For the LSD data, we have quoted the values from Table III and IV in Ref.~\cite{LatticeStrongDynamics:2018hun}.
In the combined datasets, the stability of $g_{\rho\pi\pi}$ against mass becomes more pronounced.
In the overlapped $\rho$-mass range, we find small differences
which may stem from the use of different lattice spacings.

{
In KSRF-II, our results $g_{\rho\pi\pi}\simeq 5.7$ originates
from the spectral data $M_\rho/(F_\pi/\sqrt{2}) \simeq 8.1$.
As reported in Fig.~15 of Ref.~\cite{LatKMI:2025kti}, we have found similar ratios for $N_f=4$ and $12$.
For more precise discussions, see Refs.~\cite{LatKMI:2025kti} as well as~\cite{Appelquist:2018yqe,Nogradi:2019iek,Kotov:2021mgp}.
}

Our $g_{\rho\pi\pi}$ for eight-flavor QCD are comparable with the real-world QCD values:
From the experimental data for $\Gamma(\rho\rightarrow \pi\pi)$, one finds $g_{\rho\pi\pi}\simeq 6.0$.
Using the KSRF-I relation, one finds
$g_{\rho\pi\pi} = M_\rho F_\rho/(\sqrt{2} F_\pi^2) = g_\rho/(2 F_\pi^2) \simeq 7.0$
with $g_\rho=0.119~\text{GeV}^2$ determined from $\Gamma(\rho\rightarrow e^+ e^-)$ data.
This is somewhat larger than the above experimental value and our KSRF-I result.
In comparison, the KSRF-II indicates $g_{\rho\pi\pi} = M_\rho/F_\pi \simeq 5.9$ in real-world QCD
and closer to our KSRF-II result $g_{\rho\pi\pi} \simeq 5.7$.

The coupling  $g_{\rho\pi\pi}$ in the chiral limit gives
the decay width of techni-rho
into weak bosons mediated by techni-pions,
\begin{align}
\Gamma_\rho[\rho\to WW/WZ] = \frac{(g_{\rho\pi\pi}/\sqrt{N})^2}{48\pi}M_\rho\ ,\label{eq:Gam_rho}
\end{align}
where the weak boson masses have been neglected ($M_{W/Z}\ll M_\rho$)
and we are considering the isospin-triplet part in \SUN{N_f} generators.
The normalization factor $N$ depends on the electroweak charge assignments of the techni-fermions.
In the Farhi–Susskind (one-family) model~\cite{Farhi:1980xs}, we find $N = N_f/2 = 4$
by constructing an effective Lagrangian based on a hidden local symmetry~\cite{Fukano:2015hga,Kurachi:2014qma}.
If the charge is assigned to only one doublet (two from eight-flavors), then $N = 1$ follows.

Phenomenologically, our lattice results for $F_{\rho}$ and $g_{\rho\pi\pi}$
provides important inputs to BSM model building~\cite{Fukano:2015hga,BuarqueFranzosi:2015qxd}.
For the small $m_f$ region, our estimate $g_{\rho\pi\pi}\lesssim 6$
(taking account of larger uncertainty in the KSRF-I than the KSRF-II)
predicts $\Gamma_{\rho}/M_{\rho}\lesssim 0.24/N$, with $N$ depending on the choice of model as described above.
This prediction for $\Gamma_{\rho}$ could be tested in future collider physics~\cite{Murayama:2014yja}.

The couplings $g_{\rho\pi\pi}$ from KSRF-I and II are summarized
in the fourth and fifth columns in Table.~\ref{tab:ksrf_wsr}, respectively.

\subsection{Weinberg sum rules and the Das-Mathur-Okubo (DMO) sum rule}\label{subsec:wsr}

To investigate the validity of the Weinberg and DMO spectral function sum rules,
we start from the dispersion relation,
\begin{align}
\Pi(q^2) = \frac{q^2}{12\pi}\int_0^{\infty}\frac{ds}{\pi}\ \frac{R_{\rho}(s) - R_{a_1}(s)}{s + q^2} - F^2_{\pi}\ ,
\label{eq:disp}
\end{align}
where $R_{\rho}(R_{a_1})$ represents the vector(axial-vector) spectral function.
This gives the DMOSR (the ``Zeroth Weinberg SR'')~\cite{Dash:1967fq}:
\begin{align}
S= 4\pi \Pi^\prime(0)= \frac{1}{3\pi}\int_0^{\infty}ds \frac{R_{\rho}(s) - R_{a_1}(s)}{s}.
\end{align}
According to Wilson's operator product expansion (OPE),
the left-hand side of Eq.~\eqref{eq:disp} behaves in the chiral limit as
\begin{align}
\Pi(q^2) \sim \bigl(\langle \bar{\psi}\psi\rangle^{\text{ren}}\bigr)^2\frac{q^{2\gamma_m}}{q^4}\ ,\label{eq:ope}
\end{align}
where $\gamma_m =\gamma_m (\alpha(q)) \sim 0$ is a mass anomalous dimension
in the {\it asymptotically-free UV} region\footnote{
In eight-flavor QCD, $\gamma_m$ evolves to $\sim 1$
in the lower energy region where the near-conformal spectra~\cite{Aoki:2016wnc} emerge,
while $\gamma_m$ stays almost zero at UV where the OPE \eqref{eq:ope} is applicable.}.
Expanding the right-hand side of the dispersion relation (\ref{eq:disp}) in $1/q^2$,
the first two terms ($\mathcal{O}(q^0)$ and $\mathcal{O}(q^{-2})$)
do not have counterparts in Eq.~(\ref{eq:ope}).
Thus, we obtain the Weinberg sum rules:
\begin{align}
&\frac{1}{6 \pi^2} \int_0^{\infty}ds\bigl(R_{\rho}(s) - R_{a_1}(s)\bigr) =F^2_{\pi} \ ,\\
&\frac{1}{6 \pi^2} \int_0^{\infty}ds\ s\bigl(R_{\rho}(s) - R_{a_1}(s)\bigr) = 0\ .
\end{align}
In a single pole approximation without loop effects in accord with the large $N_c$ limit,
\begin{align}
R_{\rho,a_1}(s)\simeq 6 \pi^2 F_{\rho , a_1}^2\delta(s - M^2_{\rho , a_1})\ ,
\end{align}
the sum rules reduce into
\begin{align}
\text{DMOSR:} \quad &
S_{\rm DMO}= 2\pi \biggl( \frac{F_\rho^2}{M_\rho^2} - \frac{F_{a_1}^2}{M^2_{a_1}}\biggr)\ ,
\label{DMO}\\
\text{WSR-I:}\quad &F_{\rho}^2 - F_{a_1}^2 =F^2_{\pi} \ ,\label{eq:wsr1}\\
\text{WSR-II:}\quad &F_{\rho}^2 M_{\rho}^2 - F_{a_1}^2 M_{a_1}^2 = 0\ .\label{eq:wsr2}
\end{align}
The sum rules become relevant in the chiral limit, describing
the spontaneous breaking of the chiral symmetry
$F_\pi\ne 0 \, ( F_{\rho}/F_{a_1} >1), \, M_\pi =0$,
for any $N_f$ in the broken phase, with $F_{\rho}/F_{a_1} >1$
being consistent with Fig.~\ref{fig:frho} near the chiral limit.

\begin{figure}[tbh]
\includegraphics[scale=0.94]{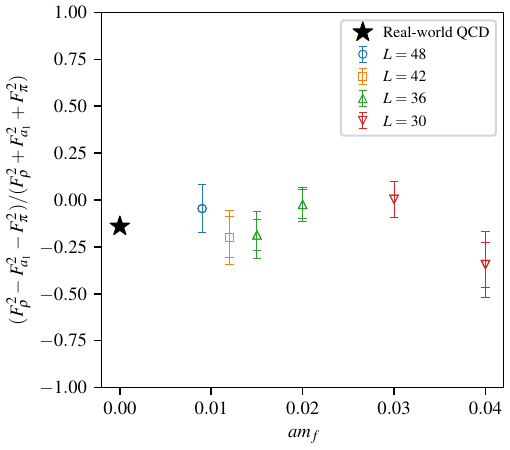}
\hfill
\includegraphics[scale=0.94]{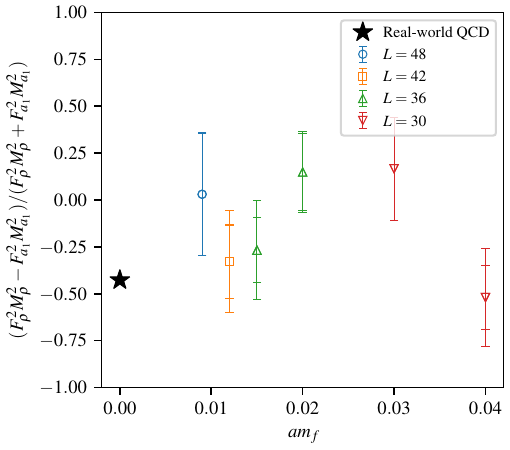}

\caption{
  The normalised WSR-I (left) and WSR-II (right), as a function of the fermion mass.
  Inner error bars are statistical, while the outer error bars are the combined statistical
  and systematic uncertainties.
}\label{fig:wsr}
\end{figure}

Figure~\ref{fig:wsr} shows the difference between the left- and right-hand sides of
the WSRs.
The results in both cases show a mild dependence on $m_f$,
and become consistent with zero within uncertainties in the small mass region.
The total errors in the WSR-I case are about 20\% or less comparing to the denominator,
and the vanishing feature is statistically significant.

For WSR-II, although the total error is larger than the WSR-I case,
the vanishing feature is still significant (50\% or less over the denominator).
Thus, we find a trend that the WSR-II is also satisfied in the small mass region
within the large uncertainties;
reducing this uncertainty is a desirable target for future work
in order to draw more precise conclusions.

\begin{table}[!tbp]
\caption{
Summary table of KSRF relations and Weinberg sum rules.
The fourth and fifth columns show Eqs.~\protect\ref{eq:ksrf1} and \protect\ref{eq:ksrf2}, respectively.
The sixth and seventh columns correspond to the normalized Weinberg sum rules I and II respectively, as plotted
in Fig.~\protect\ref{fig:wsr}.
See text for details.
The first and second parentheses show the statistical and systematic uncertainties, respectively.
}\label{tab:ksrf_wsr}
\begin{ruledtabular}
\begin{tabular}{lllllll}
\toprule
$L$ & $T$ & $am_f$ & $g_{\rho\pi\pi}^{\textnormal{\scriptsize{KSRF-I}}}$ & $g_{\rho\pi\pi}^{\textnormal{\scriptsize{KSRF-II}}}$ & WSR-I & WSR-II \\
\midrule
48 & 64 & 0.009 & $5.53(30)(25)$ & $5.628(100)(68)$ & $-0.05(13)(1)$ & $0.03(33)(4)$ \\
42 & 56 & 0.012 & $5.76(32)(25)$ & $5.708(99)(65)$ & $-0.20(11)(10)$ & $-0.33(20)(19)$ \\
36 & 48 & 0.015 & $5.52(20)(6)$ & $5.670(70)(17)$ & $-0.19(8)(10)$ & $-0.27(17)(20)$ \\
36 & 48 & 0.02 & $5.32(22)(36)$ & $5.605(71)(99)$ & $-0.02(8)(5)$ & $0.15(21)(5)$ \\
30 & 40 & 0.03 & $5.54(14)(1)$ & $5.710(45)(2)$ & $0.00(10)(0)$ & $0.17(27)(1)$ \\
30 & 40 & 0.04 & $5.29(16)(20)$ & $5.697(43)(47)$ & $-0.34(12)(13)$ & $-0.52(17)(20)$ \\
\bottomrule
\end{tabular}

\end{ruledtabular}
\end{table}

The $S$ parameter evaluated by the DMOSR, Eq.~(\ref{DMO})
can be estimated with our $F_{\rho,a_1}$ and $M_{\rho,a_1}$ shown in Table~\ref{tab:va_spectra}.
The results are shown in the sixth column of Table~\ref{tab:s_fin}.
Within large uncertainty (which originates from $F_{a_1}$ and $M_{a_1}$),
$S_{\text{DMO}}$ is roughly consistent with the VP-momentum and time-moment results.

Assuming the relation $F_\rho / F_\pi = \sqrt{2}$ (see discussions for the KSRF-I and II),
the WSR-I results in the ratio $F_\rho / F_{a_1} = \sqrt{2}$, and thereby,
the WSR-II leads to the Weinberg mass relation $M_{a_1}/M_\rho = \sqrt{2}$.
Our data are consistent with all of them, as seen in Figs.~\ref{fig:vectormass} and \ref{fig:frho},
within the uncertainty 
Using these ratios, the DMOSR $S$ parameter takes the form
\begin{align}
S_{\rm DMO}
\to S_{\rm DMO}^{\text{(reduced)}}
:= 4\pi \left(\frac{F_\pi^2}{M_\rho^2} -\frac{1}{4} \cdot \frac{F_\pi^2}{M_\rho^2} \right) 
=\frac{3\pi}{g_{\rho\pi\pi}^2} \gtrsim 0.26\ ,
\label{DMO2}
\end{align}
with our result $g_{\rho\pi\pi} \lesssim 6$ (Fig.~\ref{fig:ksrf}).
The reduced-DMOSR $S$-parameter $S_{\rm DMO}^{\text{(reduced)}}$
is roughly consistent with both VP-momentum and time-moment $S$ parameters.

\clearpage

\section{Discussion}\label{sec:discuss}

\subsection{Finite volume effects}\label{subsec:fv}

The electroweak precision test~\cite{ParticleDataGroup:2020ssz} indicates a vanishing
or at most tiny value of the $S$ parameter.
The $S$ parameter in eight-flavor QCD is implied to decrease in the small $m_f$ region
due to the approximately conformal feature~\cite{Appelquist:2014zsa}.
However, finite volume (FV) effects also suppress $S$, as we have seen in Secs.~\ref{sec:s_mom} and \ref{sec:tmr}.
In this section, we further investigate the FV effects in terms of $M_{\pi}L$ and evaluate a
{\it FV-corrected} $S$. For $M_{\pi}$, we quote the results from our independent works~\cite{LatKMI:2025kti,Aoki:2016wnc},
which are summarized in Appendix~\ref{app:spectra_ref}.

\subsubsection{$S$ parameter vs. $LM_{\pi}$}

\begin{figure}[tbh]
\includegraphics[scale=0.94]{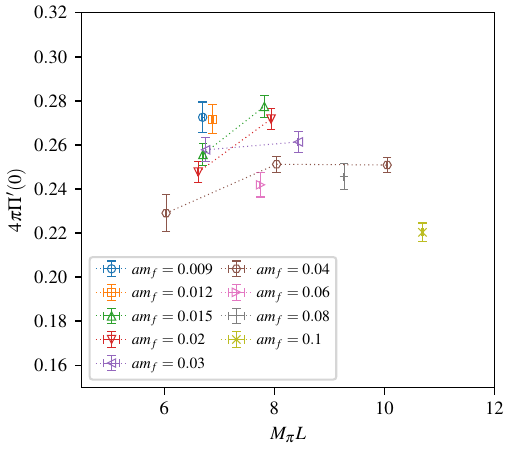}
\includegraphics[scale=0.94]{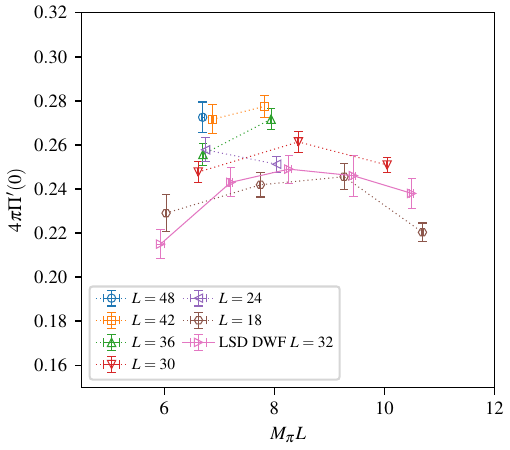}
 \caption{Strong-dynamics contribution to the $S$ parameter per
 electroweak doublet as a function of $M_\pi L$.
 The data points with the same color are obtained from the same quark mass $m_f$ (left)
 or the same volume $L$ (right).}
\label{fig:S_FV}
\end{figure}

We show the $S$ parameter calculated through the VP-momentum method in Fig.~\ref{fig:S_FV} as a function of $M_\pi L$.
Tracing the same $m_f$ (the same color symbols) in the left panel,
we find that $S$ is suppressed in the $M_{\pi}L \lesssim 8$ region.
This is interpreted as a finite volume (FV) effect, which is quite large;
for example, the suppression at $m_f=0.015$ from $L=36$ to $42$ is about 8\%,
while the pion mass shift at the same point
is only 0.04\% and statistically consistent with zero.
As shown in Appendix~\ref{app:parity_double},
the vector and axial-vector meson masses tend to move toward parity doubling
as the volume is reduced, which is qualitatively consistent
with the observed suppression of the S parameter.

In the right panel, we compare our results for $S$ with those from the LSD Collaboration~\cite{Appelquist:2014zsa}; the latter are available only for the single volume $L = 32$ (purple rightward-pointing triangles\footnote{
The LSD Collaboration’s data for the $S$ parameter have been kindly provided by David Schaich.
}).
Both the present data and the LSD Collaboration’s
show the bending-down with decreasing $M_{\pi}L$.
In our data, the decrease results solely
from FV effects as shown in the left panel.
In the LSD data, the FV was argued to be small~\cite{Appelquist:2014zsa};
however the similarity to our data implies that they may also suffer from FV.

\subsubsection{FV-corrected $S$ parameter}

We shall evaluate FV corrections to the $S$ parameter.
We adopt a fit model motivated by the next-to-leading-order staggered chiral perturbation theory (NLO-SChPT)
with one-loop two-pions contributions\footnote{
The one-loop effect is customary expressed as {\it next-to-leading} since it is an one-loop contribution.
In this terminology, there is no leading-order (LO) contribution for FV effects.}.
The model contains some minor differences from the original NLO-SChPT.
The leading FV effects result from light states wrapping around the spatial direction in the vector and axial-vector currents.
In our system, the lightest mode is an asymptotic rho-meson state in the vector current correlator:
Unlike the real-world QCD, the rho-meson mass
$M_{\rho}$ is lighter than the lowest momentum two-pion state $2\sqrt{M^2_{\pi} + (2\pi/L)^2}$.
However, the leading rho-meson correlator
creates neither loop diagrams nor FV effects.
Therefore, the second lightest contributions, two-pion intermediate states, are the leading source of FV effects.

As we have seen in Fig.~\ref{fig:S_FV}, the bending-down of $S$ 
scales with $LM_{\pi}$:
\begin{align}
S(L, M_{\pi}) = S_{\infty}(M_{\pi}) + \Delta^{\rm FV} S(LM_{\pi})\ ,\label{eq:fit_s}
\end{align}
where the left-hand side represents our lattice data, while the first and second terms in the right-hand-side
denote the infinite volume $S$ parameter and the FV correction, respectively.
We focus on the fact that the $LM_{\pi}$ scaling
appears in FV effects evaluated by NLO-SChPT with a two-pion one-loop correlator,
\begin{align}
\Delta^{\rm FV}G_{\pi\pi}(t, LM_{\pi})
 = \frac{M^3_{\pi}}{3}\sum_{n^2=1}^{\infty}\nu_n
 \int_0^{\infty} \frac{d\hat{p}\ \hat{p}^3}{2\pi^2} \frac{\sin[\hat{p}nLM_{\pi}]}{nLM_{\pi}}
 \sum_{\xi}
   \frac{e^{-2(t/L)\cdot LM_{\pi} \sqrt{\hat{p}^2 + \hat{M}^{2}_{\pi,\xi}}}}{\hat{p}^2 + \hat{M}^{2}_{\pi,\xi}}\ .
 \label{eq:XPT2pi}
\end{align}
Here, $\hat{p}$ shows pion loop momenta normalized by the NG pion mass $M_{\pi}$.
The prefactor $\nu_n = \sum_{|\vec{n}| = n}1$ is a multiplicity of pions wrapping $n$ times spatially.
The integral over $\hat{p}$ with $\nu_n$ incorporates the FV effects with a pion loop:
{
if $F(p^2)$ is an arbitrary integral function, then the FV effects on it read:}
\begin{align}
&\frac{1}{L^3}\sum_{\vec{p}_j}F(\vec{p}_j\cdot\vec{p}_j) - \int\frac{d^3p}{(2\pi)^3}F(p^2) \nn\\
&= \int\frac{d^3p}{(2\pi)^3}\sum_{n > 0}e^{i\vec{n}\cdot\vec{p}L}F(p^2)
= \sum_{n^2 = 1}^\infty\frac{\nu_n}{nL}\int\frac{dp}{2\pi^2}~p\sin[npL]F(p^2)\ ,
\end{align}
where the first equality is a Poisson summation formula.
In Eq.~(\ref{eq:XPT2pi}),
the summation over $\xi$ counts the staggered multiplet degrees of freedom:
\begin{align}
 \xi \in \{\xi_5,\ \xi_4\xi_5,\ \xi_i\xi_5,\ \xi_i\xi_4,\ \xi_i\xi_j,\ \xi_4,\ \xi_i,\ \xi_{I}\}\ ,\quad i(j) = 1,2,3\ ,
\end{align}
and $\hat{M}_{\pi,\xi}$ denotes taste multiplet pion masses normalized by NG pion mass.
In our eight-flavor system,
we have confirmed $\hat{M}_{\pi,\xi}\simeq 1$~\cite{Aoki:2016wnc}, in contrast to the usual QCD case.
Therefore, we approximate it to be unity.
The FV formula (\ref{eq:XPT2pi}) is shown and used
in the study of muon anomalous magnetic moment~\cite{Aubin:2019usy,Borsanyi:2020mff}.

Using Eq.~(\ref{eq:XPT2pi}), we construct the $S$ parameter FV correction $\Delta^{\rm FV} S(LM_{\pi})$
in a parallel way to the time-moment approach~(\ref{eq:S_tmr}),
\begin{align}
\Delta^{\rm FV} S(LM_{\pi}) = C\cdot 4\pi \sum_{t=1}^{\infty}\frac{-t^2}{2}\cdot \Delta^{\rm FV}G_{\pi\pi}(t, LM_{\pi})\ ,
\end{align}
and our fit model reads
\begin{align}
S(L, M_{\pi})
= S_{\infty}(M_{\pi}) + C\cdot 4\pi \sum_{t=1}^{\infty}\frac{-t^2}{2}\cdot \Delta^{\rm FV}G_{\pi\pi}(t, LM_{\pi})
\ .\label{eq:s_fit}
\end{align}
The fit parameters are the mass-dependent infinite-volume $S$ parameter $S_{\infty}(M_{\pi})$ and the overall factor $C$.
When the above formula is applied to the real-world QCD coupled with QED,
the prefactor is known:
$C
= -(10/9)\cdot(1/16)$.
In the present eight-flavor case, it is non-trivial to extract $C$ on the one-doublet contributions,
and hence we treat it as a fit parameter\footnote{
Our eight-flavor data contain the FV effects from axial vector current correlators with a sigma meson
as light as a NG pion. Fitting Eq.~(\ref{eq:s_fit}) to the data, the factor $C$ would be affected
by the sigma meson contribution as well.
}.

\begin{figure}[tbh]
\includegraphics[scale=0.94]{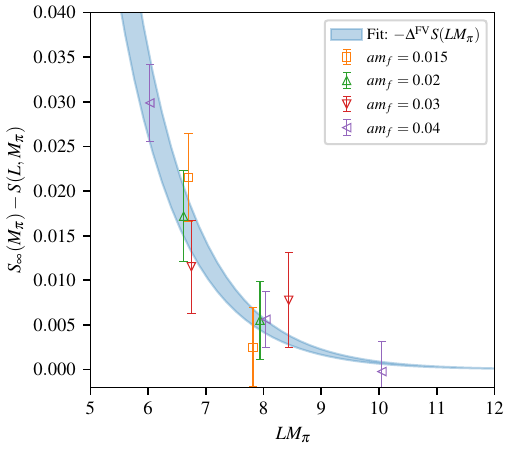}
\hfill
\includegraphics[scale=0.94]{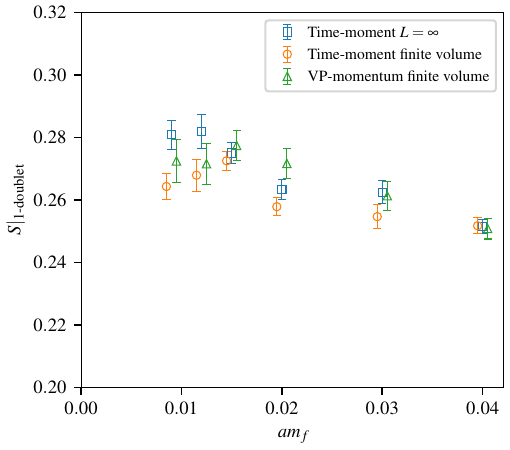}
\caption{
 Left: Finite volume correction to $S$ for a given $LM_{\pi}$.
 The light-blue band shows the fit model defined in Eq.~(\protect\ref{eq:s_fit}).
 Right: The FV-corrected $S$ (blue squares) compared with
 the original time-moment (orange circles) and VP-momentum (green triangles) results.
}
\label{fig:s_infv}
\end{figure}

Our results for the $S$ parameter have been calculated
on several different volumes for $m_f \in \MultiVolumeEnsembles$,
where the fit parameters $S_{\infty}(M_{\pi})$ and $C$ are adjusted
to explain the observed dependence of $S$ on $LM_{\pi}$.
The results are shown in the left panel of Fig.~\ref{fig:s_infv}.
The data points are the difference between the obtained $S_{\infty}(M_{\pi})$ and the original data $S(L, M_{\pi})$
and the shaded blue band is $\Delta^{\rm FV}S(LM_{\pi})$ with $C = \TimeMomentInfiniteVolumeC$.
For the other masses, $m_f \in \SingleVolumeEnsembles$,
the $S$ parameters are calculated in a single volume at each $m_f$,
then the infinite volume $S$ parameter $S_{\infty}(M_{\pi})$ is simply adjusted to be consistent with the above $C$.
As indicated in the figure,
the fit describes well the shape of the data, with $\chi^2/{\rm dof} = \TimeMomentInfiniteVolumeChisquareDof$.

Theoretically, applying NLO-SChPT to our eight-flavor system might not be fully justified.
But numerically, we have demonstrated that the scaling property with $LM_{\pi}$ can be captured by the NLO-SChPT
by introducing one universal parameter $C$.
This supports our assumption that the bending-down found in Fig.~\ref{fig:S_FV} stems from FV effects.

We shall now extract our final estimate for the $S$ parameter.
As a conservative estimate, we assign the FV-systematic error from the difference between $S_{\infty}(M_{\pi})$
and the original data $S(L, M_{\pi}) = S_{\text{TM}}$ with the largest volume at each $m_f$.
The FV-systematic error is then combined with the uncertainties in $S_{\text{TM}}$ estimated in Sec.~\ref{sec:tmr}.
The final values including all systematics, which we call $S_{\text{TM},\infty}$,
are tabulated in the fourth column of Table~\ref{tab:s_fin}.

In the right panel of Fig.~\ref{fig:s_infv},
we plot the $S_{\text{TM},\infty}$ (blue squares), which are
compared with the largest volume results of the time-moment (orange circles, $S_{\text{TM}}$)
and VP-momentum (green triangles, $S_{\text{VP-Mom}}$).
At the two smallest $m_f$, the uncertainty of $S_{\text{TM},\infty}$ becomes about 7\%,
where the FV systematics dominates.
At $m_f = 0.02$, there was a two-sigma tension between $S_{\text{TM}}$ and $S_{\text{VP-Mom}}$
as explained in Sec.~\ref{sec:tmr}, and now $S_{\text{TM},\infty}$ becomes consistent to $S_{\text{VP-Mom}}$.
This indicates that FV effects in $S_{\text{VP-Mom}}$ at $m_f = 0.02$ are small
if the largest volume ($L_s = 36$) is adopted.
At $m_f = 0.04$, the FV effect is very small: our data span three volumes in this case,
and the results confirm that the largest volume is sufficient.
For heavier $m_f$, which are not shown in the figure but given in Table~\ref{tab:s_fin}, the FV effects are tiny.

\begin{figure}[tbh]
\begin{center}
\includegraphics{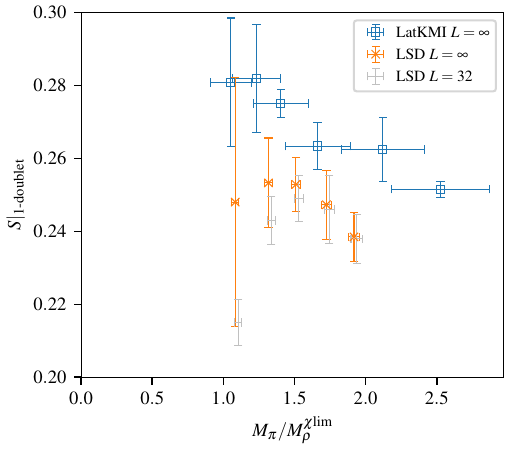}
\end{center}
\caption{
 Our final results of $S$ with all corrections (blue squares).
 For comparison, we quoted the LSD data
 with our FV correction added (orange $\times$) and without (gray $+$,~\protect\cite{Appelquist:2014zsa}).
}
\label{fig:s_fv}
\end{figure}

In Fig.~\ref{fig:s_fv},
we compare our $S$ parameter including all uncertainties (blue squares, fourth column of Table \ref{tab:s_fin})
with the LSD results (orange crosses: with FV corrections, gray plus: raw data~\cite{Appelquist:2014zsa}).
The horizontal axis is the pion mass normalized by the chiral limit of the rho-meson mass,
$M_{\pi}/M^{\chi{\rm lim}}_{\rho}$.
We quoted them from Ref.~\cite{LatKMI:2025kti} for our data, and from Ref.~\cite{Appelquist:2014zsa} for the LSD data.
We have assumed that FV corrections to the LSD data is characterized by the same universal scaling factor
as ours, {\it i.e.} $C = \TimeMomentInfiniteVolumeC$ in Eq.~(\ref{eq:s_fit}).
With $LM_{\pi}$ quoted from Ref.~\cite{Appelquist:2014zsa}, one can estimate FV effects (\ref{eq:s_fit}) in the LSD data. 
Strictly speaking, the factor $C$ might depend on the lattice spacing ($a$),
which is different between our system and the LSD\footnote{
If we estimate the lattice spacing ratio from the matching of the chiral limit rho-meson mass,
we find $a_{\rm LSD}/a_{\rm KMI} \simeq \LatticeSpacingRatioLSDLatKMI$.}.
However, FV effects associated with two-pion loops are subject to infra-red dynamics and insensitive to
ultra-violet (UV, lattice spacing) artifacts\footnote{
As explained in the remark after Eq.~(\protect\ref{eq:XPT2pi}),
the taste symmetry breaking is tiny in our eight-flavor system.
This is different to what is observed in studies of the muon $g-2$~\cite{Borsanyi:2020mff}.},
the lattice spacing effects on the FV corrections are negligible in the present case.
FV systematic errors to the LSD data are given so as to cover the central values
of the original data as we have done for our $S$.
Our results and the FV corrected LSD data become closer to each other.
The remaining difference can be interpreted as lattice spacing effects
on the original $S$ (without FV).

Our FV-corrected data for $S$ and their LSD counterparts
show a similar $M_{\pi}$ dependence, and do not bend down in the small mass region.
Thus, there is no clear evidence of suppressed $S$ within the currently accessible mass region.

\begin{table}[!tbp]
\caption{
The summary table of the FV-corrected $S$ parameter $S_{\text{TM},\infty}$
and the low energy constant $L^r_{10}(\mu = M_{\rho})$ 
to be discussed in subsection~\protect\ref{GLparameter}.
The first and second bracket show the statistical and systematic uncertainty, respectively.
We also list the results with the DMOSR $S$ parameter Eq.~(\protect\ref{DMO})
and the corresponding low energy constant Eq.~(\protect\ref{eq:L10DMO}).
See text for details.
}\label{tab:s_fin}
\begin{ruledtabular}
\begin{tabular}{lllllll}
\toprule
$L$ & $T$ & $am_f$ & $S_{\mathrm{TM},\infty}$ & $L_{10}^r(M_\rho)\cdot{10}^3$ & $S_{\mathrm{DMO}}$ & $L_{10}^r(M_\rho)|_{\mathrm{DMO}}\cdot{10}^3$ \\
\midrule
48 & 64 & 0.009 & $0.281(5)(17)$ & $-5.73(11)(33)$ & $0.247(43)(5)$ & $-5.04(86)(6)$ \\
42 & 56 & 0.012 & $0.282(5)(14)$ & $-5.87(12)(29)$ & $0.193(46)(38)$ & $-4.00(91)(81)$ \\
42 & 56 & 0.015 & $0.275(3)(2)$ & $-5.82(3)(3)$ & --- & --- \\
36 & 48 & 0.015 & $0.275(3)(23)$ & $-5.84(4)(49)$ & $0.185(31)(37)$ & $-3.98(63)(74)$ \\
36 & 48 & 0.02 & $0.263(3)(6)$ & $-5.75(3)(10)$ & $0.242(27)(22)$ & $-5.25(52)(37)$ \\
30 & 40 & 0.02 & $0.263(3)(17)$ & $-5.64(4)(34)$ & --- & --- \\
30 & 40 & 0.03 & $0.262(4)(8)$ & $-5.77(5)(15)$ & $0.254(29)(1)$ & $-5.59(58)(2)$ \\
24 & 32 & 0.03 & $0.262(4)(11)$ & $-5.73(5)(21)$ & --- & --- \\
30 & 40 & 0.04 & $0.252(2)(0)$ & $-5.67(2)(0)$ & $0.105(65)(67)$ & $-2.77(1.29)(1.37)$ \\
24 & 32 & 0.04 & $0.252(2)(6)$ & $-5.70(3)(11)$ & --- & --- \\
18 & 24 & 0.04 & $0.252(2)(30)$ & $-5.79(2)(58)$ & --- & --- \\
18 & 24 & 0.06 & $0.244(3)(7)$ & $-5.64(7)(14)$ & --- & --- \\
18 & 24 & 0.08 & $0.235(2)(2)$ & $-5.61(5)(4)$ & --- & --- \\
18 & 24 & 0.1 & $0.221(3)(0)$ & $-5.41(7)(0)$ & --- & --- \\
\bottomrule
\end{tabular}

\end{ruledtabular}
\end{table}

\subsection{$S$ parameter vs $L_{10}^r(M_\rho)$}\label{GLparameter}

We investigate $L_{10}^{r}$, one of the low energy constants in the chiral perturbation theory (ChPT).
We extract it via the relation:
\begin{align}
S = -16 \pi \, {\bar L}_{10} = -16\pi 
 \biggl[
  L^r_{10}(\mu) + \frac{1}{192\pi^2}\frac{N_f}{2}\Bigl(\log\Bigl[\frac{M^2_{\pi}}{\mu^2}\Bigr] + 1\Bigr)
 \biggr]
\ .\label{eq:l10}
\end{align}
The left-hand side is our one-doublet $S$ parameter with $N_f = 8$ sea quarks.
In the right-hand side, the factor $N_f/2 = 4$ in front of the logarithm term accounts for the loop effects
from all $N_f/2$ doublets giving rise to $N_f^2 - 1$ NG pions
with degenerate mass $M_{\pi}$.
The properties of the NG bosons in the BSM phenomenology will be discussed in the next subsection.

For the left-hand side of Eq.~(\ref{eq:l10}), we use $S_{\text{TM},\infty}$ shown in Table~\ref{tab:s_fin}.
For $M_{\pi}$ in the right-hand-side, we quote the results obtained
in our independent works~\cite{Aoki:2016wnc,LatKMI:2025kti}.
The scale $\mu$ is identified with the rho meson mass
determined at a given fermion mass $m_f$~\cite{Aoki:2016wnc,LatKMI:2025kti}.
They are summarized in Appendix~\ref{app:spectra_ref}.
Our $L^r_{10}$ reads
\begin{align}
L^r_{10}(M_{\rho})
= \frac{-S}{16\pi} - \frac{1}{192\pi^2}\frac{N_f}{2}\Bigl(\log\Bigl[\frac{M^2_{\pi}}{M_{\rho}^2}\Bigr] + 1\Bigr)
\ .\label{eq:l10_ours}
\end{align}
The definition of $L^r_{10}(M_{\rho})$ is slightly different from the usual lattice QCD counterpart
which is a fit parameter determined via a ChPT extrapolation.
In the present eight-flavor system, we do not perform the ChPT analyses,
since our data are not in the chiral-logarithm regime.
Alternatively, we investigate the $M_{\pi}$ dependence of the $L^r_{10}(M_{\rho})$ based on Eq.~(\ref{eq:l10_ours}).

\begin{figure}[tbh]
\begin{center}
\includegraphics{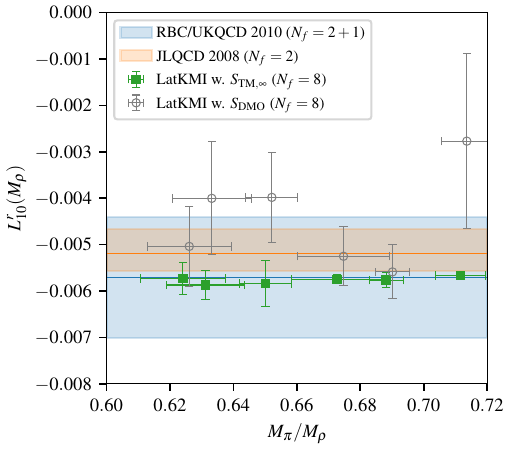}
\end{center}
\caption{
 The renormalized low energy constant $L^r_{10}(\mu = M_{\rho})$ (Table~\protect\ref{tab:s_fin}).
 The green square symbols are extracted from Eq.~(\protect\ref{eq:l10}) with
 our $S$ parameter including all corrections. 
 The gray circles are the data in Table~\protect\ref{tab:s_fin}
 obtained from $L^r_{10}(M_\rho)|_{\text{DMO}} = -S|_{\text{DMO}}/(16 \pi)$
 with the DMOSR-based $S$ parameter,
 Eq.~(\protect\ref{DMO}), using our $\rho$ and $a_1$ data.  
 For the horizontal axis, see Table~\protect\ref{tab:spectra_ref} in Appendix~\protect\ref{app:spectra_ref}.
 Our results are compared with $N_f = 2$ results
 from JLQCD group (orange band)~\protect\cite{Shintani:2008qe}
 and $N_f = 2 + 1$ results from RBC-UKQCD group~\protect\cite{Boyle:2009xi} (blue band).
 See text for details.
}
\label{fig:l10r}
\end{figure}

In Fig.~\ref{fig:l10r}, we display the $L^r_{10}(M_{\rho})$
as a function of $M_{\pi}/M_{\rho}$ (quoted from Table~\protect\ref{tab:spectra_ref} in Appendix~\protect\ref{app:spectra_ref}).
The green square symbols show the results using $S_{TM,\infty}$ in Eq.~(\ref{eq:l10_ours})
and the corresponding numbers are tabulated in the fifth column of Table~\ref{tab:s_fin}.
Our $L^r_{10}(M_{\rho})$ is almost constant against $M_{\pi}$:
\begin{align}
L^r_{10}(M_{\rho}) = -0.00573(12)(33)\quad {\rm for}\quad  m_f=0.009\ ,\label{L10r}
\end{align}
and similar values for larger $m_f$.
This observation comes from the fact that
the ratio $M_{\pi}/{M_{\rho}}$ does not vary so much, $\sim \MpiOverMrhoRange$,
and the coefficient in front of the logarithm (the pion loop effect) is small.

Next, we compare our $L^r_{10}$ with the real-world QCD value.
To clarify the relation between our definition (\ref{eq:l10_ours}) and the usual one,
we insert the chiral limit value of
the rho-meson mass $M^{\chi\text{lim}}_{\rho}$ in Eq.~(\ref{eq:l10_ours})
\begin{align}
L^r_{10}(M_{\rho})|_{M_{\pi}}
= \frac{-S(M_{\pi})}{16\pi}
 - \frac{1}{192\pi^2}\frac{N_f}{2}\biggl(\log\biggl[\frac{M^2_{\pi}}{(M^{\chi\text{lim}}_{\rho})^2}\biggr] + 1\biggr)
 - \frac{1}{192\pi^2}\frac{N_f}{2}\log\biggl[\frac{(M^{\chi\text{lim}}_{\rho})^2}{M_{\rho}^2}\biggr]\ .
\end{align}
The left-hand side depends on both $M_{\rho}$ and $M_{\pi}$, where the former dependence
is from the choice of the renormalization scale $\mu = M_{\rho}$ and accounted with the last term in the right-hand side.
Therefore,
the first and second terms in the right-hand side depend on $M_{\pi}$ only,
with a fixed $M^{\chi\text{lim}}_{\rho}$.
Approaching the chiral limit, the last term of the right-hand side would be vanishing, while
the second term with $M_{\pi}\to 0$ would show a logarithmic divergence.
The same amount of divergence should emerge from the $S$ parameter (the first term),
so that the left-hand side
\begin{align}
L^{r,\chi\text{lim}}_{10}(M^{\chi\text{lim}}_{\rho}) := L^r_{10}(M_{\rho}\to M^{\chi\text{lim}}_{\rho})|_{M_{\pi}\to 0}\ ,
\end{align}
remains finite, which becomes essentially the usual one defined in the ChPT in the real-world QCD
at physical masses $M^{\text{phys}}_{\rho} \simeq 5.5M^{\text{phys}}_{\pi} (\simeq M^{\chi\text{lim}}_{\rho})$.

The data for $L^r_{10}(M_{\rho})|_{M_{\pi}}$ we have shown so far
might be different from the chiral limit value $L^{r,\chi\text{lim}}_{10}(M^{\chi\text{lim}}_{\rho})$.
However, it is still interesting that
our results {\it coincide} with the $N_f = 2$ QCD result
at the physical rho-meson mass $M_\rho^{\text{phys}}$,
determined with hadron vacuum polarizations 
by JLQCD collaboration (shown with the orange band in Fig.~\ref{fig:l10r})~\cite{Shintani:2008qe},
$L^r_{10}(M_\rho^{\text{phys}})|_{M_{\pi}^{\text{phys}}} = - 0.0052(2)$
and $N_f = 2 + 1$ results from RBC-UKQCD collaboration~\cite{Boyle:2009xi} (the blue band in Fig.~\ref{fig:l10r}),
$L^r_{10}(M_\rho^{\text{phys}})|_{M_{\pi}^{\text{phys}}} = - 0.0057(11)$,
and also with the experimental value of the real-world QCD,
$L^r_{10}(M_\rho^{\text{phys}})|_{M_{\pi}^{\text{phys}}} = -0.0057(2)$~\cite{Harada:2003jx}
or $-0.0055(7)$~\cite{Bijnens:1994qh}.
{
Our $\bar{L}_{10}$ and $L^r_{10}$ may also be compared with the value extracted
from the DMO Eq.~(\ref{DMO}) with the QCD experimental results:
$\bar{L}_{10}|_{DMO} = - (1/8) (F_\rho^2/M_\rho^2 - F_{a_1}^2/M_{a_1}^2) \simeq -0.0077$
and $L_{10}^r(M_\rho)|_{DMO,exp} \simeq  -0.0064$,
where $F_\rho/\sqrt{2} \simeq  154$ MeV is from $\Gamma(\rho \to e^+ e^-) = 6.9(3)$ MeV,
and $F_{a_1}/\sqrt{2} \simeq 122$ MeV from $\Gamma(a_1 \to \pi \gamma) = 640(246)$ keV and $M_{a_1}=1260$ MeV.
}

{
{
To understand the coincidence mentioned above,}
we shall focus on the pion-loop contribution to $S$ (the logarithmic term in Eq.~\eqref{eq:l10}).
In eight-flavor QCD, this effect is negligible, $S^{\rm loop}_{N_f = 8}\sim 0$.
In real-world QCD with $N_f = 2 + 1$ flavors, the counterpart gives a sizable contribution}
\begin{align}
S^{\rm loop}_{\rm QCD}
= \frac{-1}{12\pi}\Bigl(
  \log\Bigl[\frac{M^2_{\pi}}{\mu^2}\Bigr] +\frac{1}{2}\log\Bigl[\frac{M^2_{K}}{\mu^2}\Bigr] + 1
 \Bigr)
\sim 0.07\ ,
\end{align}
using the experimental values of $M_{\pi,K,\rho}$.
This mostly accounts for the larger $S$ parameter observed in real-world QCD ($\sim 0.35$) in our data ($\sim 0.28$).
Thus, the remaining term $\propto L_{10}^r$ in Eq.~\eqref{eq:l10}
turns out to be similar in both theories.

We also estimate $L^r_{10}$ using DMOSR $S$ parameter (\ref{DMO})
substituted into the first term of Eq.~(\ref{eq:l10_ours}).
This corresponds to the large $N_c$ limit and has no pion-loop effects by definition.
Therefore the logarithm term in Eq.~(\ref{eq:l10_ours}) should be irrelevant:
\begin{align}
 L^r_{10}(M_\rho)|_{\text{DMO}} = \bar{L}_{10}(M_\rho)|_{\text{DMO}} = -\frac{S_{\text{DMO}}}{16 \pi}\ ,\label{eq:L10DMO}
\end{align}
which is then expected to be consistent with the result Eq.~(\ref{L10r}),
with $S^{\rm loop}_{N_f = 8} \sim 0$ at our data points.
The results are shown by the gray-circle symbols in Fig.~\ref{fig:l10r}
and the right-most column in Table~\ref{tab:s_fin}.
The errors become larger in this method due to uncertainty of $F_{a_1}$ and $M_{a_1}$.
In the smaller mass region, the two methods (the green squares and gray circles in Fig.~\ref{fig:l10r})
tend to become consistent.
If we adopt the reduced-DMOSR $S$-parameter (\ref{DMO2}) and substitute it into
the right-hand side of Eq.~(\ref{eq:L10DMO}),
we find
\begin{align}
L^r_{10}(M_\rho)|_{\text{reduced-DMO}}  \lesssim - 0.0052\ .
\label{L10rDMO}
\end{align}
The reduced-DMOSR is subject to
our observation for the decay constants ($F_{\rho}/F_{\pi}\simeq \sqrt{2},\ g_{\rho\pi\pi}\lesssim 6$)
and the assumptions that WSRs (with KSRF relations) are satisfied as discussed before Eq.~(\ref{DMO2}).
The result, Eq.~(\ref{L10rDMO}), roughly reproduces our full $S$ parameter,
which are listed in the fifth column of Table~\ref{tab:s_fin}).

In Ref.~\cite{Harada:2005ru},
the chiral limit value of the $S$ parameter has been estimated
by using the improved-ladder SD equation near the lower edge of the conformal window,
with the walking coupling close to the genuine (non-running) ladder coupling.
The setup is in accord with the anti-Veneziano limit~\cite{Matsuzaki:2015sya},
defined as $N_c\rightarrow \infty$ with $N_f/N_c\gg 1$ being fixed.
The corresponding $S$ has no pion loop effects and stays finite in the chiral limit ($\chi$lim):
$S^{\chi\text{lim}}\simeq 0.25$ to $0.30$~\cite{Harada:2005ru}. This gives
\begin{align}
 L^{r,\chi\text{lim}}_{10}(M^{\chi\text{lim}}_{\rho})
 = -\frac{S|_{\text{DMO}}}{16 \pi}
 = -\frac{S^{\chi\text{lim}}}{16 \pi} \simeq -0.0050\quad \text{to}\quad -0.0060\ ,
\end{align}
which is consistent with our lattice data (albeit with the chiral limit not yet taken)
as well as the real-world QCD with $L^r_{10}(M_\rho^{\text{phys}})|_{M_\pi^{\text{phys}}}$.
{
This experience implies that
our observations, $L_{10}^r|_{N_f = 8}\sim L_{10}^r|_{N_f = 2 (+1)}$,
may also capture the properties of the chiral limit, at least qualitatively.
To assess this expectation, we need a future study with the datasets extended to a much smaller mass region.
}

\subsection{Phenomenology}
We now discuss the phenomenological implications of our results on the $S$ parameter.
The so-called
 \PTSprm\cite{Peskin:1991sw}, $S^{\rm PT}$,
for fitting the electroweak data is defined by deviation from the SM, i.e.,
subtracting the SM Higgs contributions from the full $N_D$ doublets value $N_D S$:
\begin{align}
 S^{\rm PT}= N_D S - \Delta S_{\rm SM}
 = N_D S +  \frac{1}{12 \pi} \left[\ln \frac{M_\pi^2}{M_H^2} +\frac{5}{6}\right].
\end{align}

In technicolor model building,
only the 3 NG techni-pions (one-doublet contributions from $N_f/2 = 4$)
absorbed into the $W/Z$ bosons should be taken to the chiral limit.
The IR divergence of the techni-pion loop is considered only for these 3 NG pions
and cancelled by the SM Higgs loop in the Landau gauge.
The rest $(N_f^2 - 1) - 3 = 60$ NG pions remain massive on order $\mathcal{O}$(TeV) comparative to other bound states,
by the explicit chiral symmetry breaking due to SM and ETC gauge interactions,
where the latter are enhanced by the large anomalous dimension for $N_f=8$~\cite{Matsuzaki:2015sya,Hill:2002ap}.

If we simply regard our $\sigma$ as the Higgs with our data $M_\sigma\simeq M_\pi$,
such that $M_\pi/M_H \simeq 1$, we have $-\Delta S_{\rm SM} \simeq 0.02$, 
which implies $S^{\mathrm{PT}}\approx 0.30$ in the one-doublet case ($N_D = 1$), 
and $\approx 1.14$ in the four-doublet (Fahri–Susskind) model case  ($N_D = 4$).
The one-doublet result is somewhat smaller than the simple scale-up of the real-world QCD~\cite{Boyle:2009xi}.
However, near the chiral limit, while $M_\pi$ tends to zero,
$M_\sigma$ does not, due to the trace anomaly. As such $M_\sigma \approx M_\pi$ no longer holds,
and the estimated value will not be stable as the chiral limit is approached.

Alternatively, as an estimate more stable approaching the chiral limit, we have 
\begin{align}
- \Delta S_{\rm SM} &=\frac{1}{12 \pi} \left[ \ln
\left(
\frac{M_\pi^2}{(F_\pi^{\chi\text{lim}})^2}\cdot \frac{(F_\pi^{\chi\text{lim}})^2}{M_H^2} 
\right)
+\frac{5}{6}\right] \nonumber \\
&\sim 0.18\ (N_D=1)\ ,  \quad 0.14\ (N_D=4)\ ,\label{eq:SM_subtract}
\end{align}
with $M_\pi=0.13950(56)$ at the smallest $m_f=0.009$
in Table \ref{tab:spectra_ref} and $F_\pi^{\chi\text{lim}}=0.0210(6)$~\cite{LatKMI:2025kti},
and the identification
$F_\pi^{\chi\text{lim}}=\sqrt{2} v_{EW}=2\sqrt{2} M_H\, (N_D=1), \,\, = v_{EW}/\sqrt{2}=\sqrt{2} M_H\, (N_D=4)$.
This would give
\begin{align}
S^{\rm{PT}} \gtrsim 
\begin{cases}
0.28 + 0.18= 0.46 & (N_D=1)\ , \\
1.12 + 0.14=1.26 & (N_D=4)\ . 
\end{cases}
\label{PT-S3}
\end{align}
The logarithmic divergence in $\Delta S_{\rm SM}$ in the chiral limit is exactly
canceled by that in the $S$ (one-doublet contribution), leaving the difference staying finite.

Or, we may use Eq.~(\ref{eq:l10}):
\begin{align}
S^{\rm{PT}}
 &= -16 \pi
  \left[
    N_D L_{10}^r(M_\rho) - \frac{1}{192 \pi^2} \left(\ln \left(\frac{M_\rho^2}{M_H^2}\right)-\frac{1}{6}\right)
  \right] \nonumber \\
 &+ \frac{N_D (N_f/2 - 1)}{12 \pi} \left(\ln \frac{M_\pi^2}{M_\rho^2} + 1\right)\ .
  \label{PT-S}
\end{align}

The last term is from the contributions of the pseudo-NG pions.
In model building these do not become massless even in the chiral limit,
due to explicit breaking by electroweak and extended technicolor (ETC) interactions.
Hence, they could be regarded as similar to our data for $M_\pi^2/M_\rho^2$,
and thus negligible in both the one-doublet model ($N_D=1$) and Fahri–Susskind model ($N_D=4$).
Equation~(\ref{PT-S}) is free from the ``scale up'' estimate of the TC spectrum
with respect to $M_\pi$ vs $M_H$, and depends on more stable quantities of the lattice data.
Then our $N_f=8$ data  with $M_\rho = \RhoMassNfEightmfZeroPointZeroZeroNineLFourEightTSixFour$ at smallest $m_f=0.009$ in Table~\ref{tab:spectra_ref} indicate
$
M_\rho/M_H= (M_\rho/F_\pi^{\chi\text{lim}})/(M_H/F_\pi^{\chi\text{lim}}) \sim 30
 (N_D=1, F_\pi^{\chi\text{lim}}/\sqrt{2}= v_{\rm EW}=2 M_H)
$,
$\sim 15 (N_D=4,\ F_\pi^{\chi\text{lim}}/\sqrt{2} = v_{\rm EW}/2=M_H)$,
almost independently of $m_f$.
Substituting them into Eq.~(\ref{PT-S}), we find a result consistent with Eq.~(\ref{PT-S3}).

Our result for $N_D = 1$ (the first line in Eq.~(\ref{PT-S3})) may be compared with
the naive scale-up of real-world QCD~\cite{Appelquist:2014zsa}
\begin{align}
S^{\rm{PT}} = S_{\rm QCD} -\Delta S_{\rm SM} \simeq 0.35 + 0.08 = 0.43\qquad (N_D=1)\ .
\end{align}
For the first term contribution,
the eight-flavor result $S\simeq 0.28$ is smaller than the real world counterpart $S_{\rm QCD}\simeq 0.35$.
For the second term effect,
a key quantity is the ratio $M_\pi/F_\pi^{\chi\text{lim}}$, which is $\gtrsim 6.6$ in the eight-flavor case
and larger than that in the real world $M_\pi/F_\pi^{\chi\text{lim}}\simeq 1.1$.
This is the reason why the eight-flavor result $-\Delta S_{\rm SM} \gtrsim 0.18$
becomes larger than the real world counterpart $0.08$.
As a result, the total in the eight-flavor case $S^{\rm PT} \gtrsim 0.46$ is somewhat larger than the real world value $0.43$.
In view of Eq.~(\ref{PT-S}),
the $L_{10}^{r}(M_\rho)$ term is almost universal independently of $N_f$
while the logarithmic term in the first line is somewhat larger in the present theory ($\simeq 0.18$)
comparing to that in real-world QCD ($\simeq 0.14$).

{
The original motivation of the present paper is to investigate whether the $S$ parameter can be reduced
within the pure WTC which is realized by the $N_f=8$ QCD, even without effects of the ETC and SM gauge interactions.
In fact, our result ($S\simeq \SParameterInfiniteVolumeTimeMomentNfEightmfZeroPointZeroZeroNine$ for $m_f = 0.009$) is smaller than the scaled-up version of $N_f = 2+1$-flavor QCD
$S \simeq 0.35$. However, the Peskin-Takeuchi $S$ parameter $S^{\rm PT} = N_D S - \Delta S_{\rm SM}$
is not suppressed as indicated in Eq.~(\ref{PT-S3}).
This does not necessarily rule out the WTC model itself~\cite{Matsuzaki:2015sya}, 
since the contributions from the TC sector can easily be cancelled
by the strong mixing with the SM fermion contribution through the ETC interactions~\cite{Dimopoulos:1979es},
as in the fermion delocalization of the Higgsless model~\cite{Cacciapaglia:2004rb}.
The ETC is needed for the realistic model building to give mass to the quarks and leptons.
This was actually the very starting point of
the WTC with large anomalous dimension to solve the FCNC problem.
Thus the $S$ parameter to be compared with the electroweak data should be calculated
not just within the WTC in isolation but in the whole ETC dynamics.
}

\clearpage

\section{Summary and Outlook}\label{sec:summary}

We have investigated the Peskin-Takeuchi
$S$ parameter~\cite{Peskin:1991sw} using lattice gauge theory
to assess the proposal that the $SU(3)$ gauge theory with eight-flavor QCD
causes electroweak symmetry breaking.
Our numerical results have indicated the eight-flavor QCD
to be just below the lower edge of the conformal window along the $N_f$ axis
and an approximately conformal (walking) theory with large mass anomalous dimension $\gamma_m \sim 1$.
The eight-flavor QCD thus provides a candidate basis theory for the walking technicolor (WTC)
model~\cite{Yamawaki:1985zg,Bando:1986bg,Holdom:1984sk,Akiba:1985rr,Appelquist:1986an},
where the walking dynamics could resolve various problems known in technicolor models.
The $S$ parameter is one such problem, and has here been studied using first-principles computations.

We have used the same dataset as the comprehensive previous studies on the eight-flavor hadron spectra~\cite{Aoki:2016wnc}
and added one ensemble with the lightest fermion mass $m_f = 0.009$~\cite{LatKMI:2025kti}.
We have used the tree-level Symanzik-improved gauge action~\cite{Follana:2006rc,MILC:2010pul},
and the highly improved staggered quarks (HISQ) action~\cite{Symanzik:1983dc}.
No fourth root is needed in the latter, as the two species of staggered fermion correspond
to eight flavours with exact chiral symmetry, up to taste symmetry breaking.
Our HISQ ensembles with $N_f = 8$ have been confirmed to acquire
at most 5\% taste violation even in the pion channel~\cite{Aoki:2016wnc}.
Thus, our lattice setup has been well-tested and suited for the $S$ parameter study.

We have measured the vector and axial-vector current correlators using the $N_f = 8$ gauge configurations.
We have applied two methods to $S$ parameter calculations.
The first one (VP-Momentum method) is the conventional approach
where the $S$ parameter has been extracted from the zero-momentum slope of the transverse component ($\Pi(q^2)$)
of the vacuum polarization (VP),
{\it i.e.}~the Fourier transformation of the difference of the two correlators ($V-A$).
The second method utilizes so-called time momentum representation~\cite{Bernecker:2011gh}
and the \Sprm is directly evaluated in the coordinate space from the first time moment of the $V-A$ correlators (TM method).
This work is the first application of the TM method to the computation of the $S$ parameter.

In the VP-Momentum method, Pad\'{e} fits for $\Pi(q^2)$ have been performed with $q_{\text{min}} = 2\pi/T$.
For the maximum momenta, we have taken all possible cases with $|q_{i = 1,2,3}| = n\times 2\pi/L$
where $n = 1$ for $L = 18$ and $n = 2$ for $L \in \{24,30,36,42,48\}$.
The fit has worked in all ensembles, with $\chi^2/{\text{dof}}\sim 0.1 - 0.5$.
In the TM method, we have modeled the large temporal distance behavior of the $V-A$ correlator
by using the vector and axial-vector meson masses which have been precisely estimated in our spectrum paper~\cite{Aoki:2016wnc}.
This has enabled us to take the large $T$ limit analytically.
Both methods have provided the consistent results up to minor exceptions
(Fig.~\ref{fig:sparam_tmr_vs_mom} and Table~\ref{tab:s_lat}).

We have evaluate the vector and axial-vector decay constants (Table~\ref{tab:va_spectra})
and investigated the Weinberg sum-rules (WSR) and the KSRF relations.
We have found that both of WSR-I and II are satisfied.
The result for WSR-I indicates the spontaneously broken chiral symmetry in the eight-flavor QCD,
consistently to the hadron spectrum studies~\cite{Aoki:2016wnc,Aoki:2014oha,Aoki:2013xza}.

By using the decay constants, we have also investigated the KSRF-I and II.
Based on our recent works~\cite{Aoki:2016wnc,LatKMI:2025kti},
we have assumed that chiral symmetry is spontaneously broken in this theory,
and thereby, the KSRF-I holds\footnote{
A fate of the chiral symmetry in eight-flavor QCD with the chiral and continuum limits taken
is discussed in Refs~\cite{LatKMI:2025kti,Witzel:2024bly,Hasenfratz:2022qan}.
In the present work, we have assumed spontaneous breaking of chiral symmetry
based on our works using the same gauge configurations as this work.}.
This has allowed us to estimate the coupling between techni-rho and two techni-pions $g_{\rho\pi\pi}\lesssim 6$
at small mass region. The result is similar to the real-world QCD case.
Using this, we have predicted the decay width of techni-rhos into weak bosons mediated by techni-pions:
$\Gamma_{\rho}/M_{\rho}\lesssim 0.24/N$
with $N = 4$ (Farhi-Susskind model) or $N = 1$ (one-doublet charge assignment).
Our $\Gamma_{\rho}$ could be confirmed/{ruled out}
in future collider experiments~\cite{Murayama:2014yja}.

The KSRF-I combined with $F_\rho / F_\pi = \sqrt{2}$ results in the KSRF-II.
In fact, our data for the decay constant show that ratio, which has been a non-trivial observation.
Thus, the KSRF-II is also satisfied in our eight-flavor system.

Based on the above results, we have discussed two subjects:
(1) Finite volume (FV) corrections to the $S$ parameter, and 
(2) Implication of our $S$ to BSM phenomenology.

In both of VP-Momentum and TM methods, as well as the previous result by LSD collaboration~\cite{Appelquist:2014zsa},
the $S$ parameter shows a similar suppression when $M_{\pi}L$ becomes smaller than $8$.
This has suggested much stronger finite volume effects than those observed in the hadron spectra~\cite{Aoki:2016wnc}.
In order to correct the FV effects,
we have utilized the next-to-leading-order (NLO)
staggered chiral perturbation theory (SChPT) where the aforementioned $M_{\pi}L$ scaling naturally appears
from two-pions wrapping around the spatial directions~\cite{Borsanyi:2020mff,Aubin:2019usy}.
In our $V-A$ current correlators, such two pions emerge in the vector part.
We have not directly applied the SChPT to our data but construct a fit model
by introducing two fit parameters: the infinite volume $S$ parameter and the prefactor to the SChPT expression.
This fit has worked, giving $\chi^2/{\text{dof}}\sim 1$.
Conservatively, we have assigned the FV systematic errors
so that they covers the largest volume data point at each $m_f$.
The FV-corrected $S$ parameters are summarized in the sixth column in Table~\ref{tab:s_fin}
and have not shown any clear trend of suppression at small $m_f$.

{
Our FV-corrected $S$ parameter at the smallest mass $\SParameterInfiniteVolumeTimeMomentNfEightmfZeroPointZeroZeroNine$ is somewhat smaller than the real-world QCD case
($S\sim 0.35(3)$~\cite{Harada:2003jx}, without SM subtraction).
Interestingly, the Gasser-Leutwyler parameter obtained with our $S$ parameter data (see Eq.~(\ref{eq:l10_ours}))
is almost independent of pion masses and the result at the smallest mass point
$L_{10}^{r} (M_\rho) \simeq \lTenrNfEightmfZeroPointZeroZeroNineLFourEightTSixFour$ (Table~\ref{tab:s_fin})
is consistent with the value observed in experiments for real-world QCD,
$L_{10}^{r} (M_\rho) \simeq -0.0055(2)$},
which is of direct relevance to the Peskin-Takeuchi $S$ parameter with subtraction of the SM Higgs contributions
to be compared with the  electroweak experiments.
This gives the Peskin-Takeuchi $S$ parameter larger
than the experimental constraint, Eq.~\eqref{PT-S};
the walking (near-conformal) feature by itself does not reconcile the $S$ parameter with the experiments
within the available datasets.
This is opposed to our original expectation that a small $S$ might emerge thanks to the near-conformality,
while does not rule out the WTC model itself~\cite{Matsuzaki:2015sya};
the $S$ parameter to be compared with the electroweak data results from the whole ETC dynamics
where the sizable WTC contribution can easily be cancelled by the SM fermion contribution
as we discussed in the previous section.

As a future perspective, the $S$ parameter, KSRF relations, Weinberg sum rules, and the low energy constant
should be further investigated by using the datasets with smaller masses and taking the chiral extrapolation.
The study of the ETC effects on the lattice will also be of important issue.
We believe that the present work has provided a firm step to attack those future subjects.

\begin{acknowledgments}
We thank the Lattice Strong Dynamics (LSD) collaboration for sharing their plots and their preliminary unpublished results for the flavor-singlet scalar mass on their ensembles. 
We gratefully acknowledge Dr.~Masafumi Kurachi and Dr.~Kei-ichi Nagai 
or their collaboration during the early stages of this work.
Numerical calculations have been carried out on the high-performance computing systems at KMI ({\Large$\varphi$}),
at the Information Technology Center in Nagoya University (CX400),
and at the Research Institute for Information Technology in Kyushu University (CX400 and HA8000)
both through the HPCI System Research Projects (Project ID: hp140152, hp150157, hp160153) and through general use.
This work is supported by the JSPS Grants-in-Aid for Scientific Research (S) No. 22224003, (C) No. 16K05320 (Y.A.)
for Young Scientists (A) No.16H06002 (T.Y.), (B) No.25800138 (T.Y.), (B) No.25800139 (H.O.), (B) No.15K17644 (K.M.). 
E.B.~acknowledges the support of
the UKRI Science and Technology Facilities Council (STFC) Research Software Engineering Fellowship EP/V052489/1,
the EPSRC ExCALIBUR programme ExaTEPP (project EP/X017168/1),
the STFC Consolidated Grant No.\ ST/T000813/1, and the Supercomputing Wales programme,
which is part-funded by the European Regional Development Fund (ERDF) via Welsh Government.
K.M. is supported in part by the JSPS KAKENHI (No.~26K21728).
H.O. is supported in part by the JSPS KAKENHI (Nos.~21K03554, 22H00138).
T.Y. is supported in part by Grants-in-Aid 
for Scientific Research (Nos.~19H01892, 23H01195, 23K25891) and
MEXT as ``Program for Promoting Researches on the Supercomputer Fugaku''
Grant Number JPMXP1020230409.
This work is supported by the JLDG constructed over the SINET6 of NII.
This work was in part based on the MILC collaboration’s public lattice gauge theory code.
See \url{http://physics.utah.edu/~detar/milc.html}.
\end{acknowledgments}

\section*{Open access}
For the purpose of open access, the authors have applied a Creative Commons Attribution (CC BY) licence to any author accepted manuscript version arising.

\section*{Data availability statement}
Raw and processed data generated during the preparation of this work, and the analysis workflows to transform these data into the form presented in this work, are available at Refs.~\cite{datapackage,workflow}.

\clearpage

\appendix
\section{HISQ currents}\label{app:hisq_current}

Let us first recall the vector and axial vector currents
in the naive staggered fermions composed with anti $\phi^{(i)}$ and $\phi^{(j)}$ staggered fermion fields,
\begin{eqnarray}
 \Vc^{(i,j)}_\mu(x) & = & 
  \frac{1}{2}\left\{\phibar^{(i)}(x+\muhat)\eta_\mu(x) U^\dag_\mu(x)\phi^{(j)}(x)
  + \phibar^{(i)}(x)\eta_\mu(x) U_\mu(x)\phi^{(j)}(x+\muhat)\right\}\ ,
 \label{eqn:Vmu}\\
 \Ac^{(i,j)}_\mu(x) & = & 
  \frac{1}{2}\left\{\phibar^{(i)}(x+\muhat)\eta_\mu(x)\epsilon(x) U^\dag_\mu(x)\phi^{(j)}(x)
  - \phibar^{(i)}(x)\eta_\mu(x)\epsilon(x) U_\mu(x)\phi^{(j)}(x+\muhat)\right\}\ ,
 \label{eqn:Amu}
\end{eqnarray}
where the usual phase factors are assumed (
$\eta_\mu(x)=(-1)^{\sum_{\nu<\mu}x_\nu}$, 
$\epsilon(x)=(-1)^{\sum_{\mu}x_\mu}$
).
$j\ne i$. $i$ and $j$ take $1$ or $2$ for the eight flavor theory.
These currents satisfy the ``conservation'' laws
\begin{eqnarray}
 \Delta^-_\mu \Vc^{(i,j)}_\mu(x) & = & 0\ , \label{eq:cclV}\\
 \Delta^-_\mu \Ac^{(i,j)}_\mu(x) & = & 2m_q \phibar^{(i)}(x)\epsilon(x)\phi^{(j)}(x)\ ,\label{eq:cclA}
\end{eqnarray}
where $\Delta^-$ is the backward difference operator.
These currents and also those for the HISQ formulation have the spin-taste structure,
\begin{equation}
 \begin{array}{cc} 
 (\gamma_\mu\otimes 1), & (\gamma_\mu\gamma_5\otimes\xi_5)\ .\\
 \end{array}
\end{equation}
Before defining the HISQ currents, let us introduce the following notation for bilinear point-split operators with length $n$,
\begin{eqnarray}
 \Kmu{n}{Y}(x) & = & \phibar^{(i)}(x+n\muhat) \eta_\mu(x) \{
  Y^{\dagger}_\mu(x+(n-1)\muhat) \cdots
  Y^{\dagger}_\mu(x+\muhat) \cdot
  Y^{\dagger}_\mu(x)\}
  \phi^{(j)}(x)\ ,\\ \nonumber
  & - & \phibar^{(i)}(x) \eta_\mu(x) \{
  Y_\mu(x)\cdot 
  Y_\mu(x+\muhat)
  \cdots
  Y_\mu(x+(n-1)\muhat)\}
  \phi^{(j)}(x+n\muhat),\\
 \Ldmu{n}{Y}(x)& = & \phibar^{(i)}(x)\eta_\mu(x)\epsilon(x) \{
  Y_\mu(x)\cdot 
  Y_\mu(x+\muhat)
  \cdots
  Y_\mu(x+(n-1)\muhat)\}
  \phi^{(j)}(x+n\muhat),\\ \nonumber
  & - & \phibar^{(i)}(x+n\muhat)\eta_\mu(x)\epsilon(x) \{
  Y^{\dagger}_\mu(x+(n-1)\muhat) \cdots
  Y^{\dagger}_\mu(x+\muhat) \cdot
  Y^{\dagger}_\mu(x)\}
  \phi^{(j)}(x)\ ,
\end{eqnarray}
where $Y_\mu(x)$ indicate a link field deformed from $U_\mu(x)$ with a certain smearing procedure.
The $i$ and $j$ indices are suppressed in the left hand side but assumed.
Using this notation the conserved currents of the naive staggered fermions
Eqs.~(\ref{eqn:Vmu}) and (\ref{eqn:Amu}), which are length one point-split 
currents, are written as
\begin{equation}
 \Vc_\mu^{(i,j)}(x) = \frac{1}{2} \Kmu{1}{U}, \; \Ac_\mu^{(i,j)}(x) = \frac{1}{2} \Ldmu{1}{U}\ .
\end{equation}
Now for the HISQ \cite{Follana:2006rc,MILC:2010pul}, we have two types of smeared link
$W_\mu(x)$ and $X_\mu(x)$, which enter the definition of the conserved currents.
They read
\begin{eqnarray}
 W_\mu(x) & = & {\mathcal P}\ {\mathcal U}^\mathrm{fat7} U_\mu(x)\ ,\\
 \label{eq:Wlink}
 X_\mu(x) & = & {\mathcal U}^\mathrm{fat7}\ W_\mu(x)
  + 2\ {\mathcal U}^\mathrm{Lepage}\ W_\mu(x)\ ,
 \label{eq:Xlink}
\end{eqnarray}
where the right hand side of the first equation means first to apply
fat7 smearing on the original link $U_\mu$ and then to take 
the unitary projection to one of the $U(3)$ elements \cite{MILC:2010pul}.
The second term in the last equation is the Lepage term made of $W_\mu(x)$,
twice of the Lepage term in the asqtad action. 
Detailed decomposition of these are conveniently written in Appendix A
of Ref.~\cite{MILC:2010pul}.
Using these definitions the conserved currents for HISQ are written as
\begin{eqnarray}
 \Vc_\mu^{(i,j)}(x) & = & 
  \frac{1}{2} \Kmu{1}{X}(x) 
 - \frac{1}{48} \left\{ (\Kmu{3}{W}(x) + \Kmu{3}{W}(x-\muhat) + \Kmu{3}{W}(x-2\muhat)) 
   \right\}
 \label{eq:hisq_ccv}\\
 \Ac_\mu^{(i,j)}(x) & = & 
  \frac{1}{2} \Ldmu{1}{X}(x)
  - \frac{1}{48} \left\{ (\Ldmu{3}{W}(x) + \Ldmu{3}{W}(x-\muhat) + \Ldmu{3}{W}(x-2\muhat))
   \right\}\ .
 \label{eq:hisq_cca}
\end{eqnarray}
The length-one terms contain $X_\mu$ links in Eq.~\eqref{eq:Xlink},
while the length-three terms have only $W_\mu$ links in Eq.~\eqref{eq:Wlink}.
One can obtain these currents by an usual variational method with infinitesimal 
vector or axial transformation of $\overline{\phi}^{(i)}$ and $\phi^{(j)}$.
These currents satisfy the conservation laws, Eqs.~(\ref{eq:cclV}) and
(\ref{eq:cclA}), with the HISQ action.

The vacuum polarization function which matches the continuum definition,
Eq.~(\ref{eq:Pi_mn}), reads
\begin{equation}
 \Pi_{\mu\nu}(q) = \frac{1}{4}\sum_{x}e^{iqx}
  \left\{
   \langle \Vc^{(i,j)}_\mu(x) \Vc^{(j,i)}_\nu(0)\rangle
   - \langle \Ac^{(i,j)}_\mu(x) \Ac^{(j,i)}_\nu(0)\rangle
  \right\}\ ,
\end{equation}
where the prefactor $1/4$ is due to the staggered taste multiplicity.
For practical reasons we use the conserved currents for sink ($x$ position)
and one-link currents which have the same spin-taste structure as the conserved
currents for the source (at the origin),
\begin{equation}
 \Pi_{\mu\nu}(q) = \frac{1}{4}\sum_{x}e^{iqx}
  \left\{
   Z_V \langle \Vc^{(i,j)}_\mu(x) V^{(j,i)}_\nu(0)\rangle
   - Z_A \langle \Ac^{(i,j)}_\mu(x) A^{(j,i)}_\nu(0)\rangle
  \right\}\ ,
\end{equation}
where $Z_V$ ($Z_A$) renormalizes $V_\mu$ ($A_\mu$), which is 
the one link currents consisting of the HISQ smeared link $X_\mu$,
\begin{equation}
 V^{(i,j)}_\mu(x) = \frac{1}{2} \Kmu{1}{X}, \; A^{(i,j)}_\mu(x) = \frac{1}{2} \Ldmu{1}{X}\ .
  \label{eq:hisq_olcs}
\end{equation}
Due to the exact chiral symmetry of the staggered fermions we can set 
$Z_V=Z_A$.
As discussed in Sec.~\ref{sec:basics}, the use of these currents ensures the 
extraction of the needed vacuum polarization function which is free from UV divergences.

\section{Fitting range of $\Pi(q^2)$}
\label{app:fit_range}

Our main results for the $S$ parameter from the vacuum polarization function
in the momentum space ($S_{\text{VP-Mom}}$) is obtained by fitting to a Pad\'e form~\eqref{eq:pade}
with the data in a momentum range.
Here we investigates the fit-range systematics.

The smallest possible $q^2$ is given by $q_\mu=(0,0,0,\pm 2\pi/T)$ with 
$L < T$, and here $L$ and $T$ are the spatial and temporal size of the lattice\footnote{
With $q^2=0$ (all $q_\mu=0$), the transverse part cannot be singled out.}:

\begin{align}
 (T,L) \in \AllLatticeVolumes\ .
\end{align}
Thus, the lower bound of the fit range is
\begin{align}
 q^2_{\min} = (2\pi/T)^2\ .
\end{align}
For the upper bound,
we have considered three possible momenta:
\begin{align}
 & 1\ (\text{in lattice units})\ ,\nonumber\\
 & q^2_{{\max}2} = 3\cdot (2\cdot 2\pi/L)^2 + (2\cdot 2\pi/T)^2\ ,\nonumber\\
 & q^2_{{\max}3} = 3\cdot (3\cdot 2\pi/L)^2 + (3\cdot 2\pi/T)^2\ .
\end{align}
We evaluate the $\Pi(q^2)$ with all combinations of $q_\mu$ satisfying
\begin{align}
 q^2 < q^2_{\max} \equiv \min\{q^2_{{\max}2}, 1\}\ .
\end{align}
Here, the upper bound ``1'' is selected only for the smallest volume $(T, L) = (24, 18)$.
Whether or not these choices bias the final results can be checked by
investigating the dependence with respect to the change of the range:
$1 \leftrightarrow q^2_{{\max}2}$ for the smallest volume,
$q^2_{{\max}2} \leftrightarrow q^2_{{\max}3}$ for the others.
Results obtained in this way have been reported in Sec.~\ref{sec:s_mom}.

\begin{figure}[tbh]
\includegraphics[scale=0.94]{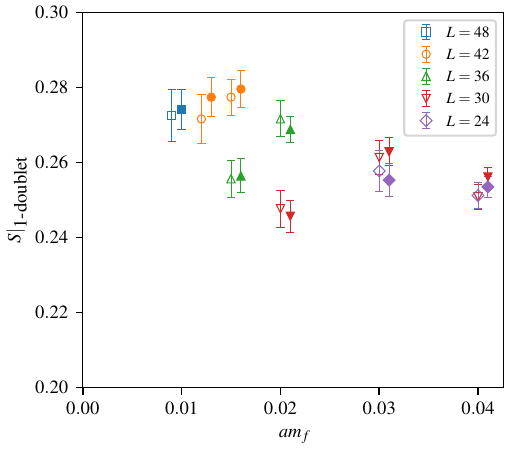}
\hfill
\includegraphics[scale=0.94]{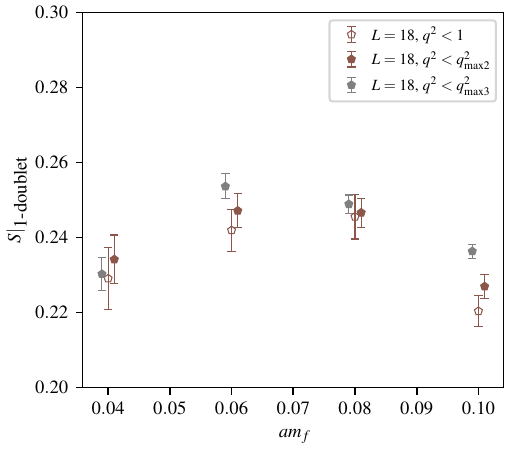}
\caption{
The $S$ parameter as a function of $m_f$ with various upper bounds of
the momentum range in the Pad\'e fits~\protect\eqref{eq:pade}.
Left: Results in the small mass region $m_f\le 0.04$.
Open and filled symbols show the results from the fits with the momentum range
satisfying $q^2 < q^2_{{\max}2}$ and $q^2 < q^2_{{\max}3}$, respectively.
Right: The smallest volume results in the large mass region $m_f \geq 0.04$.
Open brown, filled brown, and filled gray symbols show
the results from the fit with the momentum range satisfying $q^2 < 1$, $q^2 < q^2_{{\max}2}$,
and $q^2 < q^2_{{\max}3}$, respectively.
}
\label{fig:S-qmax23}
\end{figure}

Figure \ref{fig:S-qmax23} compares the $S$ parameter (one-doublet contribution $4\pi\Pi^{\prime}(0)$)
obtained by using different upper bounds in the Pad\'e fits~\protect\eqref{eq:pade}.
The left panel  focuses on the small mass range $m_f \leq 0.04$
(The smallest volume results are omitted and focused in the right figure).
In this range, 
the uncertainty arising from the choice of either $q^2_{{\max}2}$ or $q^2_{{\max}3}$
(the difference between open and filled symbols)
is less than the statistical uncertainty.
It is therefore safe to use the smaller range $q_{{\max}2}^2$ in these Pad\'e fits.
The shift of the central values of $S$ by $q^2_{{\max}2} \leftrightarrow q^2_{{\max}3}$
is considered as a systematic uncertainty
which is added to the statistical uncertainty in quadrature to get the total error.

In the right panel of Fig.~\ref{fig:S-qmax23},
the $S$ parameter with the smallest volume $L = 18$ is displayed.
In some cases, the results fluctuate beyond statistical errors,
which indicates sizable discretization errors.
We note that $|q_{\mu}| = 3\cdot 2\pi/L|_{L = 18}$ already exceeds $1$ in lattice units.
Therefore, we should restrict the fit range momenta to $q^2 < 1$.

\begin{figure}[tbh]
\includegraphics[scale=0.94]{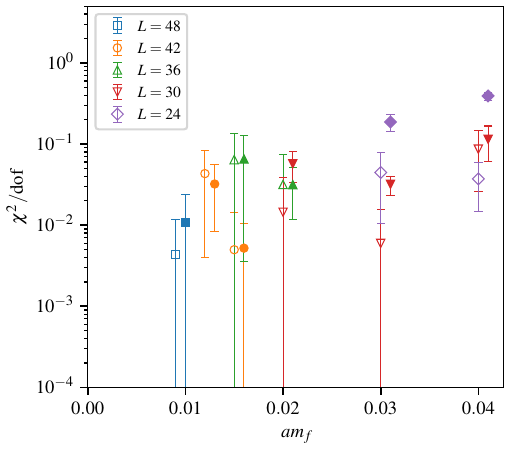}
\hfill
\includegraphics[scale=0.94]{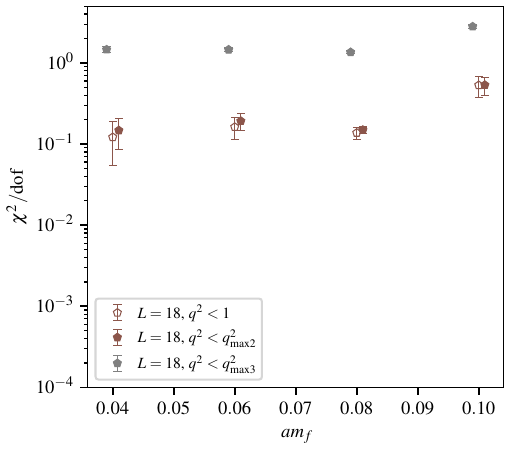}
\caption{
$\chi^2/\text{dof}$ of the Pad\'e fits~\protect\eqref{eq:pade}
for the $\Pi(q^2)$ as a function of $m_f$.
The symbol and color schemes are the same as Fig.~\protect\ref{fig:S-qmax23}.
}
\label{fig:chisq_smom}
\end{figure}

Figure~\ref{fig:chisq_smom} shows the $\chi^2/{\text{dof}}$ of the Pad\'e fits.
In the light mass region (left panel), $\chi^2/{\text{dof}}$ is small $\lesssim\mathcal{O}(1)$
and stable against the choice of the upper bounds, $q^2_{{\max}2}$ or $q^2_{{\max}3}$.
For the smallest volume results in the heavy mass region (right panel),
$\chi^2/{\text{dof}}$ becomes significantly larger when we adopt $q^2_{{\max}3}$ as the upper bound.
In this region it is less clear that $q_{\mathrm{max}3}^2$ is sufficiently small
to capture the uncertainty in the data;
this is a reflection of the discretization issue mentioned above.
In comparison, $\chi^2/{\text{dof}}$ shows a tiny dependence on the choice of $1$ or $q^2_{{\max}2}$.
This supports the reliability of the Pad\'e fits in the range $q^2 < 1$.
For the smallest volume,
the shift of the central values of $S$ by $1 \leftrightarrow q^2_{{\max}2}$
is considered as a systematic uncertainty
which is added to the statistical uncertainty in quadrature to give the total error.

\clearpage
\section{Mass spectra as reference scales}\label{app:spectra_ref}

In this Appendix, the pion decay constant $F_{\pi}$, Nambu-Goldstone pion mass $M_{\pi}$,
PV-channel vector and axial-vector meson masses $M_{\rho(PV)}$ and $M_{a_1(PV)}$ are summarized in Table~\ref{tab:spectra_ref}.
The results have been obtained in our independent works~\cite{LatKMI:2025kti,Aoki:2016wnc}
and quoted in several places in the present work as reference scales.

\begin{table}[!tbp]
\caption{
The summary table of spectra quoted in the present work.
The data with $\dagger$ are from Ref.~\protect\cite{LatKMI:2025kti} and the others are from Ref.~\protect\cite{Aoki:2016wnc}.
}\label{tab:spectra_ref}
\begin{ruledtabular}
\begin{tabular}{llllllll}
\toprule
$L$ & $T$ & $am_f$ & $aF_\pi$ & $aM_\pi$ & $aM_{\rho(\mathrm{PV})}$ & $aM_{a_1(\mathrm{PV})}$ \\
\midrule
48 & 64 & 0.009 & 0.03971(20)${}^\dagger$ & 0.13950(56)${}^\dagger$ & 0.2225(24)${}^\dagger$ & 0.3144(59)${}^\dagger$ \\
42 & 56 & 0.012 & 0.04542(27) & 0.16362(43) & 0.2536(17) & 0.346(11) \\
42 & 56 & 0.015 & 0.05054(15) & 0.18614(15) & 0.2827(21) & 0.387(10) \\
36 & 48 & 0.015 & 0.05047(14) & 0.18606(31) & 0.2815(23) & 0.3854(82) \\
36 & 48 & 0.02 & 0.05848(15) & 0.22052(33) & 0.3223(31) & 0.460(12) \\
30 & 40 & 0.02 & 0.05775(17) & 0.22232(42) & 0.3334(22) & 0.443(14) \\
30 & 40 & 0.03 & 0.07157(10) & 0.28122(24) & 0.4075(24) & 0.528(24) \\
30 & 40 & 0.04 & 0.082641(99) & 0.33501(21) & 0.4719(23) & 0.610(27) \\
24 & 32 & 0.03 & 0.07085(11) & 0.28306(34) & 0.4134(29) & 0.504(14) \\
24 & 32 & 0.04 & 0.08235(14) & 0.33487(35) & 0.4686(22) & 0.676(39) \\
18 & 24 & 0.04 & 0.09096(57) & 0.38856(15) & 0.5323(84) & 0.656(13) \\
18 & 24 & 0.06 & 0.10082(23) & 0.43170(63) & 0.5908(25) & 0.686(42) \\
18 & 24 & 0.08 & 0.11694(23) & 0.51524(50) & 0.6804(25) & 1.064(77) \\
18 & 24 & 0.1 & 0.13151(44) & 0.5948(11) & 0.7729(65) & --- \\
\bottomrule
\end{tabular}

\end{ruledtabular}
\end{table}

\section{Finite Volume and Parity Doubling}\label{app:parity_double}

In Sec.~\ref{subsec:fv}, we have discussed FV effects in the computation of the $S$ parameter on the lattice.
Related to that, we here discuss the parity doubling of the vector and axial-vector meson masses.

\begin{figure}[tbh]
\begin{center}
\includegraphics{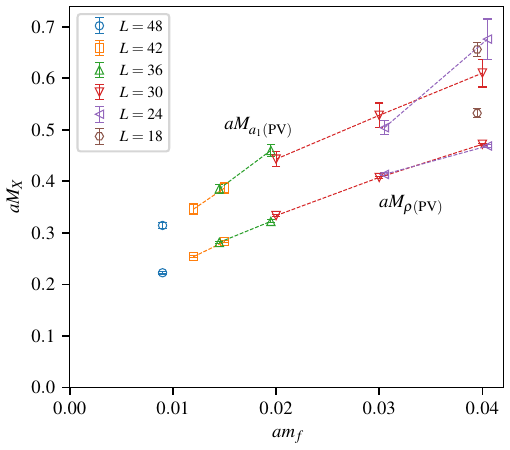}
\end{center}
\caption{
 Vector ($\rho$) and axial vector ($a_1$) meson masses measured
 with PV channel as a function of quark mass.
}
\label{fig:spectrum}
\end{figure}

We quote the mass spectra measured in the PV channel\footnote{
In Sec.~\protect\ref{sec:decay_const},
we have shown $\rho$ and $a_1$ masses ($M_{\rho,a_1}$) measured with one-link point-split operators,
which are the results of the present work and used for the studies of the KSRF relations and WSR.
However, the statistical noise in $M_{a_1}$
prevents the discussion of parity doubling,
for which the point operator results (PV channel, the sixth and seventh columns in Table~\ref{tab:spectra_ref}) 
are more suitable and quoted from Refs~\protect\cite{Aoki:2016wnc,LatKMI:2025kti}.}
$(\gamma_k\gamma_4\otimes \xi_k\xi_4)$ from Refs.~\cite{Aoki:2016wnc,LatKMI:2025kti}.
As shown in Fig.~\ref{fig:spectrum},
a remarkable FV dependence is observed for the vector meson mass, which drives the meson heavier as the volume is reduced.
In the axial vector, the response appears opposite to the vector meson, as seen at $m_f=0.02$ and 0.03.
For the other $m_f$, the trend becomes less clear due to larger statistical noise in the axial vector.
The movement to the small volume is towards the parity doubling, and thus seems consistent with the FV
effect on $S$ discussed in Sec.~\ref{subsec:fv} through the dispersive analysis using spectral decomposition~\cite{Peskin:1991sw}.
We have to note, however, the correspondence between the parity doubling and
the decrease of $S$ is not clear for several data points.
For example, $S$ decreased 8\% from $L=42$ to 36 at $m_f=0.015$,
while the meson masses at the same parameter are stable.

\bibliographystyle{apsrev4-1a}
\bibliography{LatKMI-S}

\end{document}